\documentclass[10pt,aps,twocolumn,prx,floatfix]{revtex4-2}

\usepackage{graphicx}
\usepackage{color}
\usepackage{amssymb,amsfonts,amsmath,amsthm}
\usepackage{bm}  
\usepackage{epstopdf}
\usepackage[normalem]{ulem}  

\theoremstyle{definition}   
\newtheorem*{thm*}{Theorem}

\newcommand{\hlattice}{%
  \begin{figure}[tbp]
    \centering
    \includegraphics[width=0.7\linewidth]{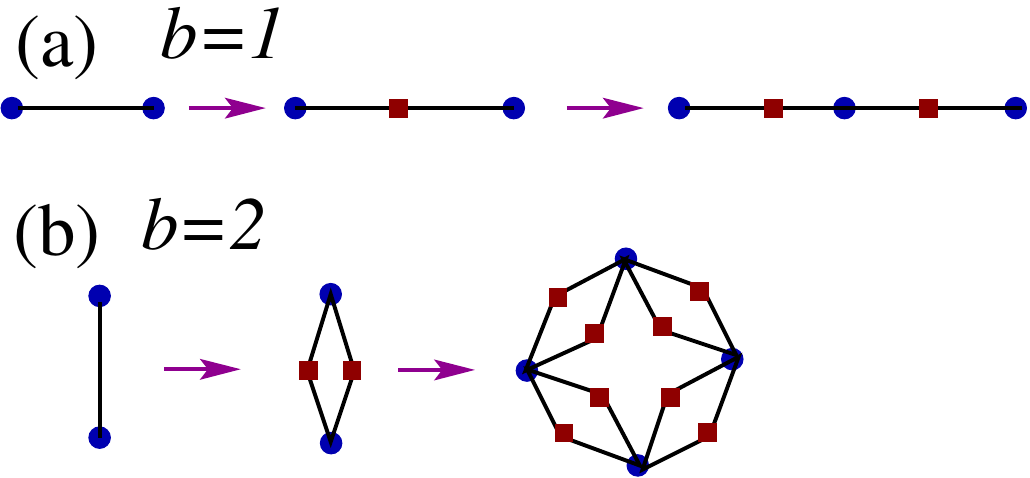} 
    \hspace{-.81cm} \includegraphics[scale=0.45]{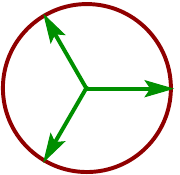}{\large{(c)}}
    \caption{Hierarchical lattices constructed by replacing a bond by
      a motif ad infinitum. Here the motif is a diamond of $b$
      branches, with each branch adding a new point (red square). Each
      motif adds $2b$ bonds. Successive lattices are indexed by the
      generation number $n$ with $n=0$ for the starting bond.  (a) A
      one dimensional lattice generated by a $b=1$ ``diamond'' (which
      is a line). (b) For $b=2$, two new sites and $2b=4$ bonds are
      created for each bond. Eventually, one gets a 2-dimensional
      lattice, though not a Bravais lattice. Three generations are
      shown in (a) and (b).  (c) A planar vector representation of a
      Potts spin with $q=3$ states. The vectors are mutually at an
      angle of $2\pi/3$}
    \label{fig:latt} \label{fig:circle}
  \end{figure}
}%

\newcommand{\perodlat}{%
  \begin{figure}[hbp]
    \centering
    \includegraphics[width=0.6\linewidth]{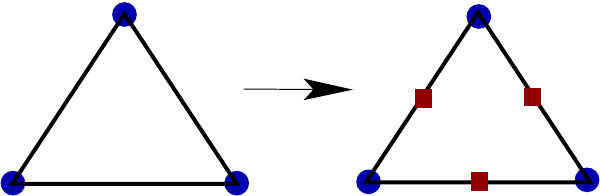} 
    \caption{A periodic chain constructed hierarchically. Each bond of
      the triangle (generation $n=0$) goes through the iteration
      process of Fig. \ref{fig:latt}a. The number of bonds for
      generation $n$ is $B_n=3\times 2^n$. }
    \label{fig:pbc}
  \end{figure}
}%

\newcommand{\figboneqthree}{%
  \begin{figure}[tbp]
    \centering
    \includegraphics[width=\linewidth]{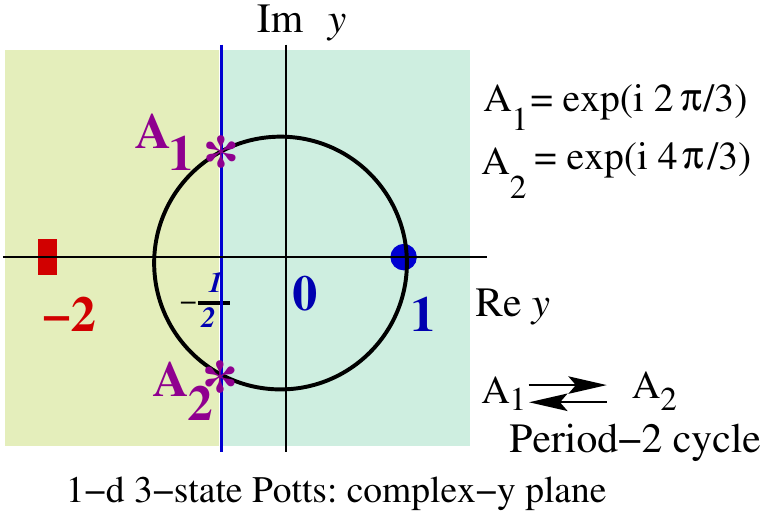}
    \caption{ RG behaviour in the complex-$y$ plane for the Potts
      chain.  Quantum time evolution takes place along the unit circle
      ($|y|=1$ or $y=e^{i\theta}$), while the thermal behaviour comes
      from the positive real axis, notably $1<y<\infty$. The blue
      vertical line at ${\rm Re } y=-1/2$ is the Julia set for the RG
      transformation $R(y)$ which has three fps, two attractive fps at
      $y^*=1$, $y^*=-2$, and a repulsive fp at $y^*=\infty$.  $A_{j},
      (j=1,2)$ represent a {\it periodic cycle of period 2}. The
      points on the arc from $A_1$ to $A_2$ on the right side (darker
      region) flow to $y^*=1$, while those on the arc on the left side
      (lighter region) of the blue line flow to $y^*=-2$.}
\label{fig:onedcircle}
  \end{figure}
}%

\newcommand{\figonedlosch}{%
  \begin{figure}[bp]
    \centering
    \includegraphics[width=\linewidth]{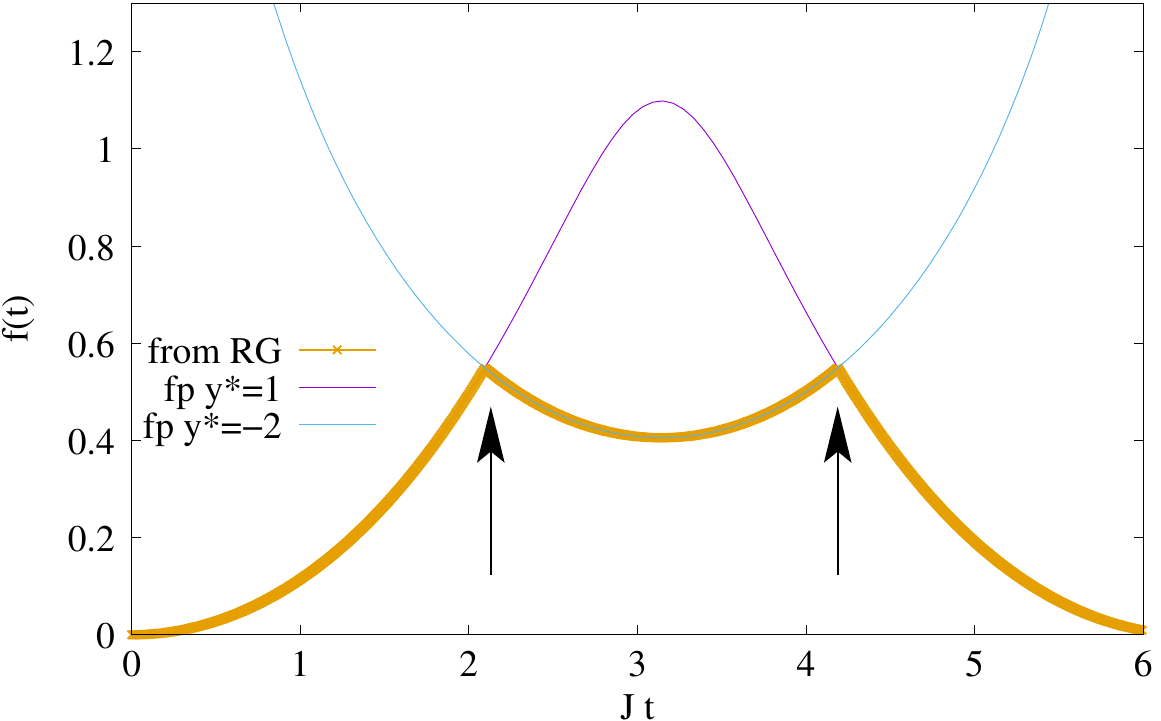}
    \caption{ The rate function vs time ($Jt$) for the periodic Potts
      chain.  The results from the sum of 60 terms (Eq. (\ref{eq:26})
      is shown by the thick line which shows two points (indicated by
      arrows) with slope discontinuities at $Jt_{c1}=2\pi/3$, and
      $Jt_{c2}=4\pi/3$. There is a periodicity---the transitions are
      seen at $Jt_{c1}+2\pi n, Jt_{c2}+2\pi n$ for any integer $n$.
      The rate functions characteristic of the two fps ($y^*=1, -2$)
      are denoted by the thin lines (Eqs.(\ref{eq:39}) and
      (\ref{eq:81})).  These overlap within numerical accuracy with
      the data points.  }%
\label{fig:fvstoned}
  \end{figure}
}%

\newcommand{\figboneopen}{%
  \begin{figure}[htbp]
    \centering
    \includegraphics[width=\linewidth]{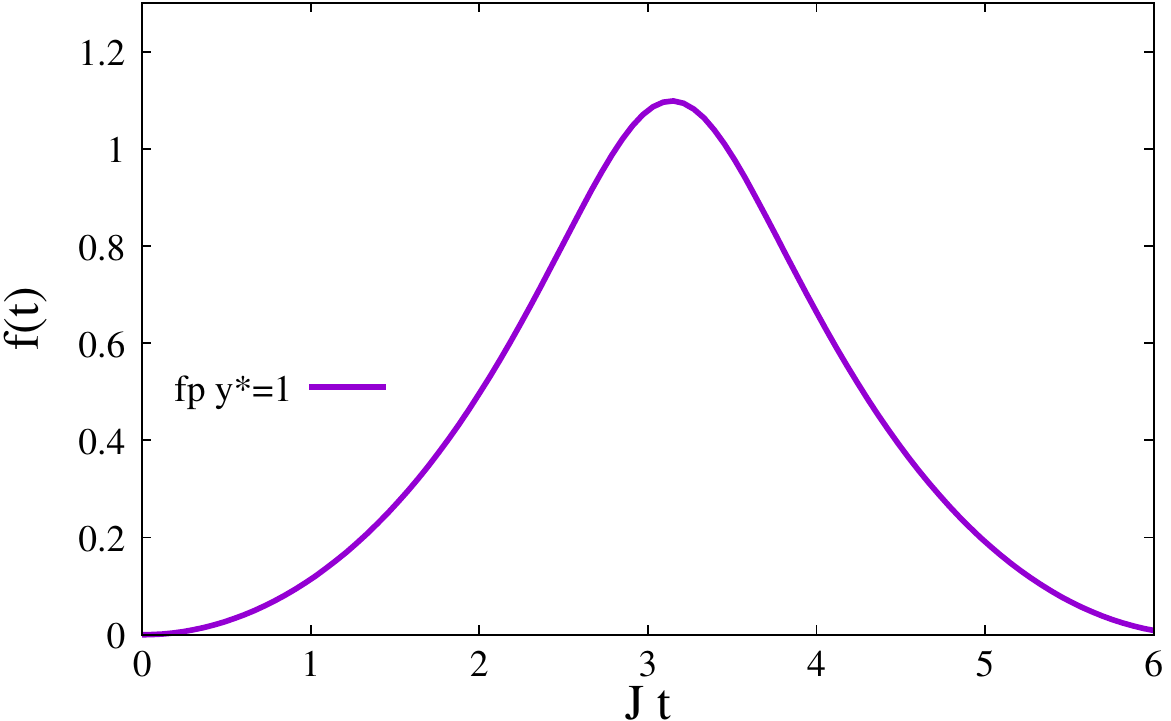}
    \caption{ Loschmidt rate function $f(t)$ vs. $t$ for the Potts
      chain under open boundary condition.  There is no DQPT but
      oscillations with periodicity $J t=2\pi$.  This function is the
      characteristic of the phase that corresponds to the fixed point
      $y^*=1$. }%
\label{fig:b1ft}
  \end{figure}
}%

\newcommand{\juliabtwo}{%
  \begin{figure}[htbp]
    \centering
  (a)
  \includegraphics[width=0.95\linewidth, trim= 10pt 20pt 80pt 110pt]{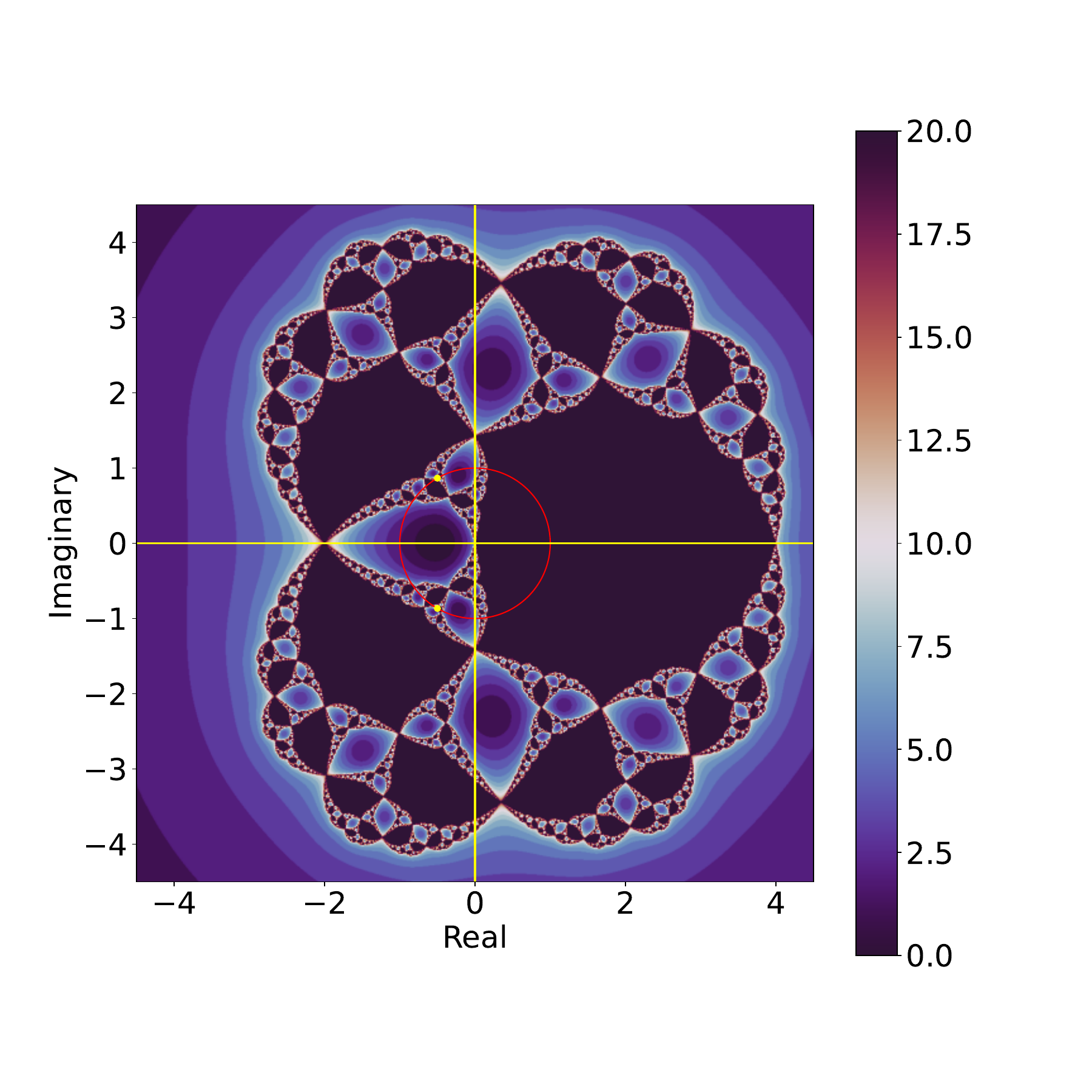}
  
  (b)  \includegraphics[width=0.95\linewidth, trim= 20pt 60pt 0 110pt]{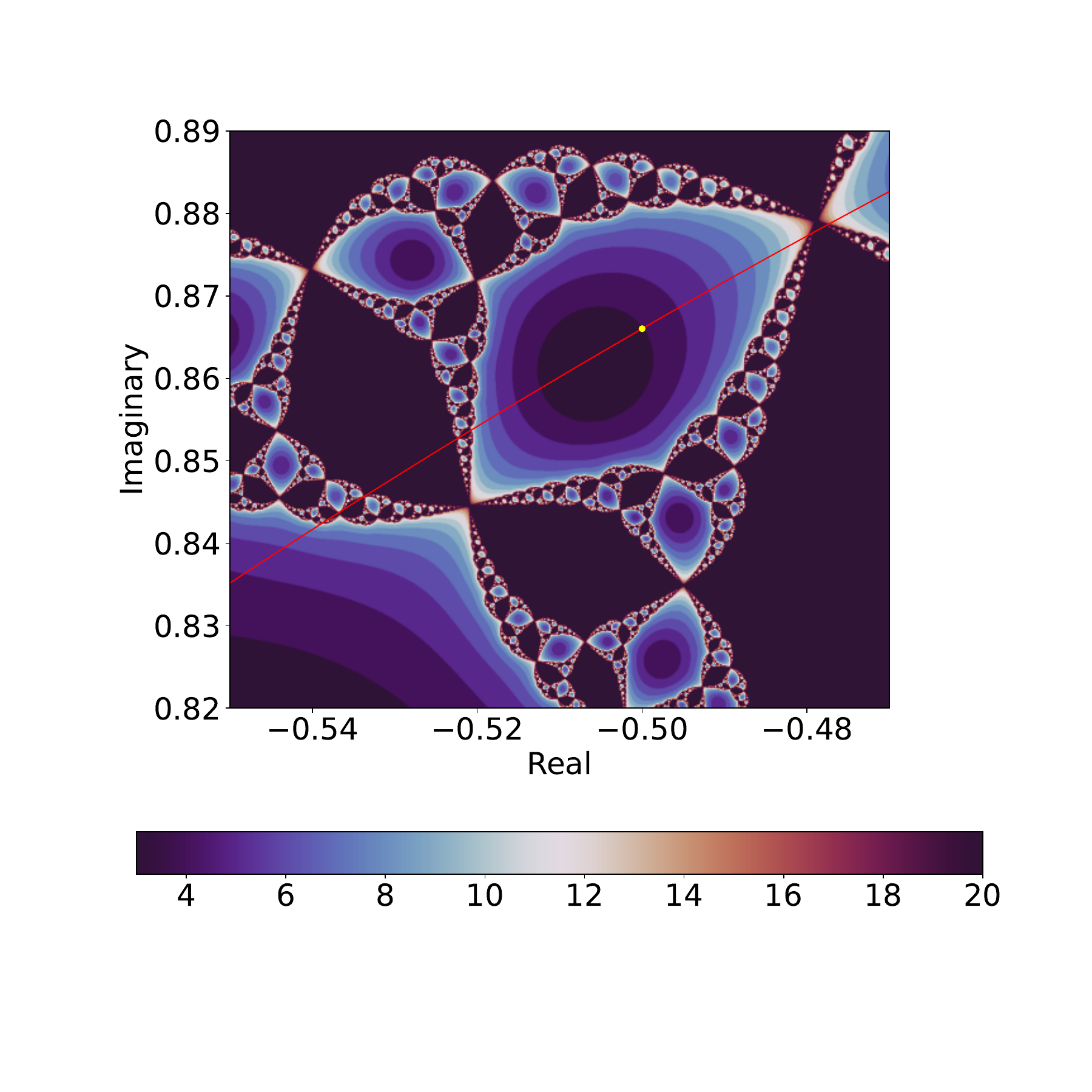}

  (c)  \includegraphics[width=0.95\linewidth, trim= 20pt 60pt 0 180pt]{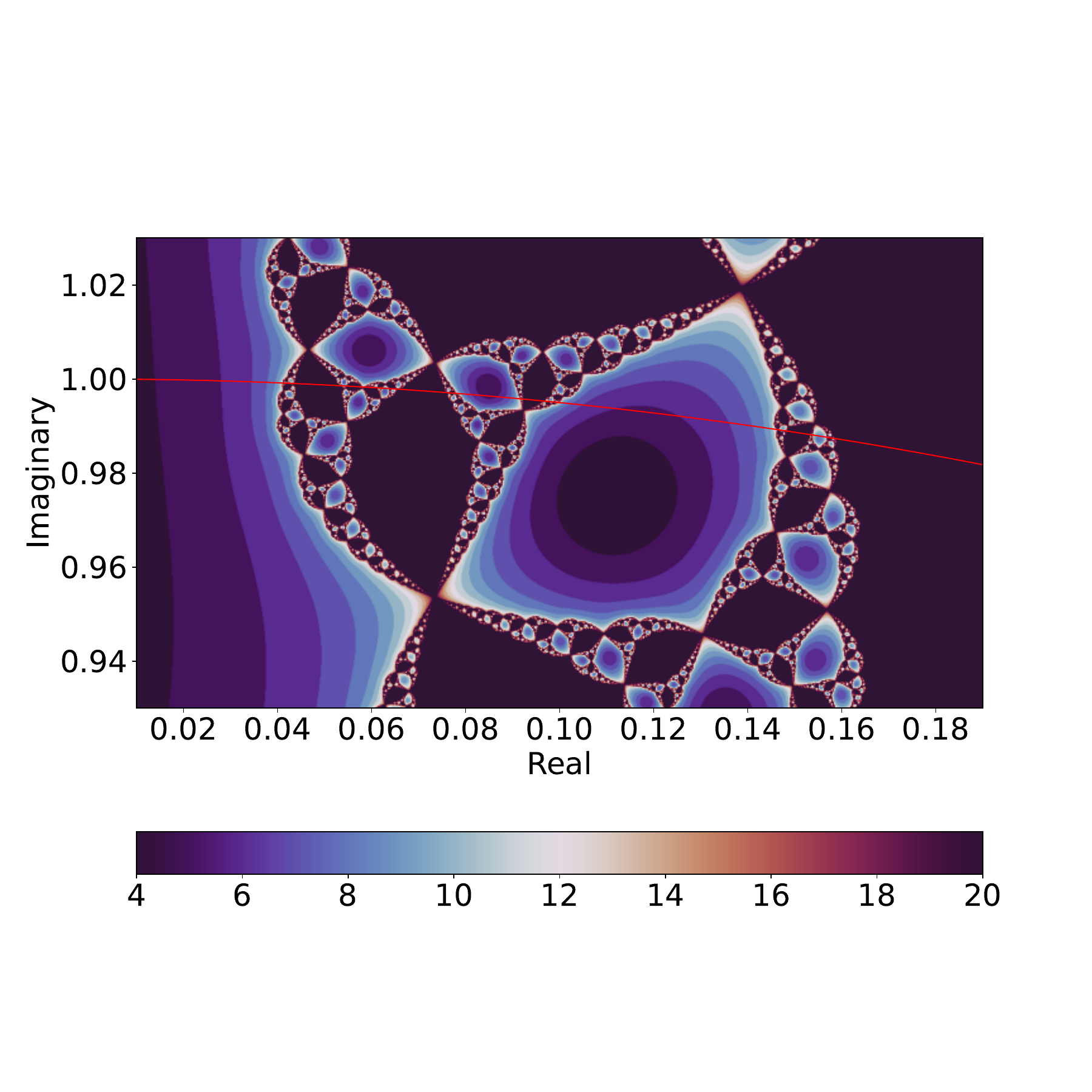}

  \caption{The Julia set for the $b=2$ RG transformation $R_2$.  The
    colour code represents the number of iterations required ($<20$)
    to reach the stable fps $y^*=1$ or $y^*=\infty$.  The complex fps
    on the unit circle are represented by yellow dots. The full Julia
    set is shown in (a) while a magnified image around the fp
    $y=\exp(i 2\pi/3)$ is shown in (b).  The intersection of the unit
    circle with the Julia set in the first quadrant is shown in (c).}
    \label{fig:b2julia}
  \end{figure}
}%

\newcommand{\btwoftvst}{%
  \begin{figure}[htbp]
    \centering
    \includegraphics[width=0.9\linewidth]{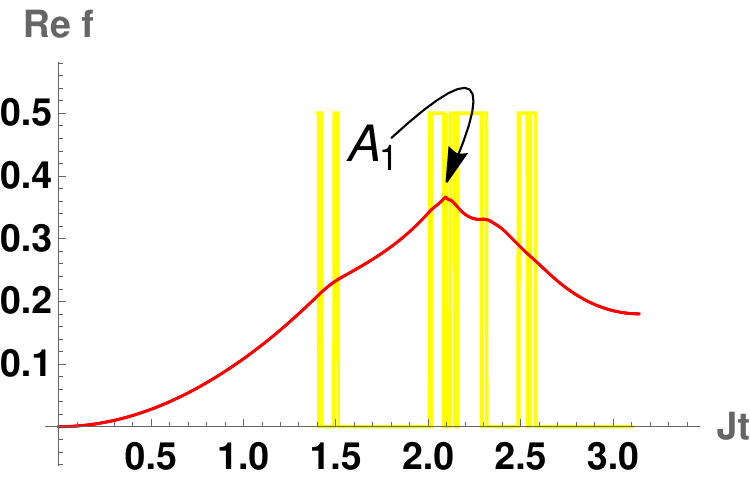}
    \caption{The rate function $f(t)$ vs $t$ (red solid curve) for
      $b=2$. The arrow marks the peak, which represents the location
      of the first-order transition at $Jt=2\pi/3=2.094...$. The
      yellow vertical lines indicate some of the zeros derived from
      the intersection of the unit circle $|y|=1$ with the Julia set.
      These singularities, which form a dense set, are not described
      by the listed fixed points but belong to higher-order periodic
      orbits.  }
   \label{fig:b2ft}
  \end{figure}
}%

\newcommand{\deltaf}{%
  \begin{figure}[htbp]
    \centering
    \includegraphics[width=0.9\linewidth]{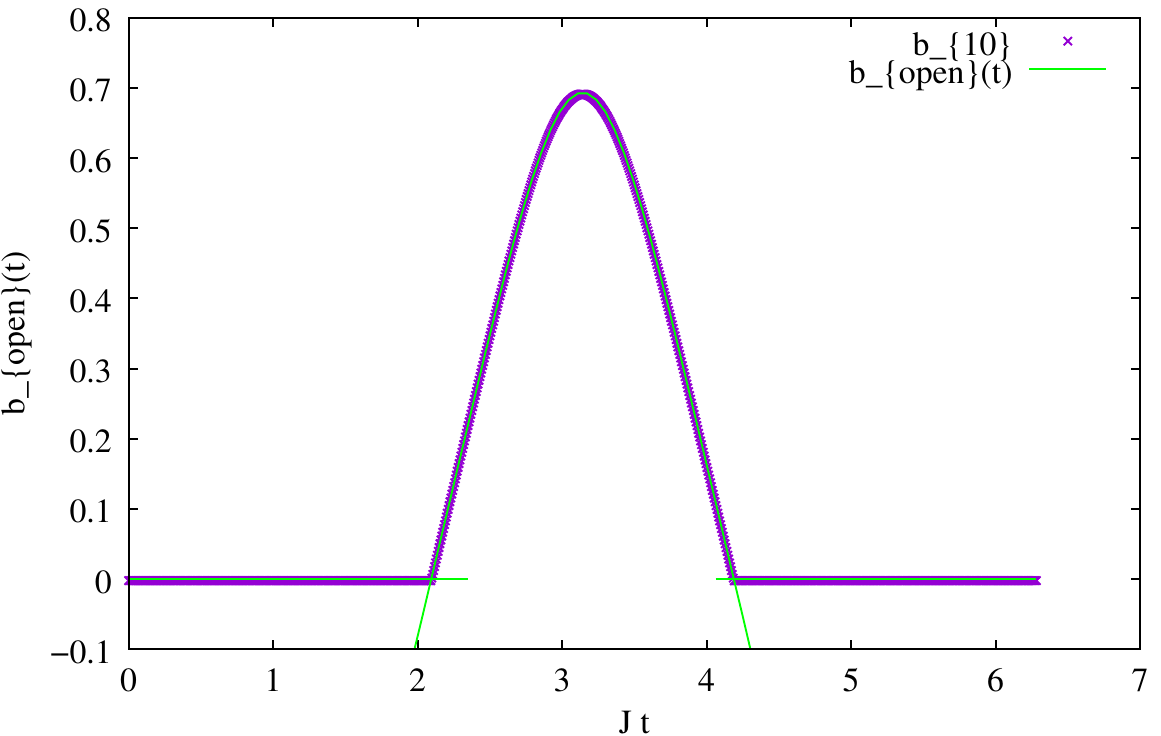}
    \caption{The remainder term $b_{open}(t)$-vs-$Jt$ for the open
      3-state Potts chain. Eqs. (\ref{eq:82}),(\ref{eq:83}) for the
      rate function (green solid line for $2\pi/3<Jt<4\pi/3$) compares
      well with the numerically evaluated values of $b_n$ for $n=10$.
      The green lines have been extended beyond the intersections for
      clarity.}
\label{fig:deltaf}
\end{figure}
}%

\newcommand{\tablebc}{%
\begin{table}[tbp]
  \centering
  \begin{tabular}{|l|c|l|c|}
\hline
\hline
Boundary & Chain type& Partition function & DQPT?\\
Conditions&  ($n=1$) &      $Z_1(y)$      &      \\
\hline
Open &  $\bullet\!\!\rule[2pt]{24pt}{1pt}\!\bullet$& $3y+6$& No\\
     &free\hspace{24pt}free&&\\
\hline
Open &  $\uparrow\!\!\!\rule[2pt]{24pt}{1pt}\!\bullet$& $y+2$& No\\
     &  fixed\hspace{24pt}free&&\\
\hline
Open &  $\uparrow\!\!\!\rule{24pt}{1pt}\!\!\!\uparrow$& $y$ (if parallel)& Yes\\
     &  fixed\hspace{24pt}fixed&1 (nonparallel)&\\
\hline
Periodic& ${{\triangle}}$ & $3 y^3+18 y + 6$& Yes\\[4pt]
\hline
Periodic&
$\bullet\!\!{\rule[3.2pt]{25pt}{1pt}}$\hspace{-23pt}\!${\rule[0.1pt]{25pt}{1pt}}\!\!\bullet$&
$3y^2+6$& Yes\\
\hline
\hline
  \end{tabular}
  \caption{Boundary conditions and DQPT.  An arrow indicates a spin of
    fixed orientation, while a bullet or a light vertex indicates a spin
    that takes all 3 possible orientations. The end point conditions 
    are preserved as the lattice is built hierarchically.}
  \label{tab:bcdqpt}
\end{table}
}%

\begin{document}

\title{Complex dynamics approach to Dynamical quantum phase
  transitions:the  Potts model}
\author{Somendra M.  Bhattacharjee}
\email{somendra.bhattacharjee\string@ashoka.edu.in}

\affiliation{Department of Physics, Ashoka University, Sonepat,  Haryana - 131029, India}

\date{\today}

\begin{abstract}
  This paper introduces complex dynamics methods to study dynamical
  quantum phase transitions in the one- and two-dimensional quantum
  3-state Potts model.  The quench involves switching off an infinite
  transverse field.  The time-dependent Loschmidt echo is evaluated by
  an exact renormalization group (RG) transformation in the complex
  plane where the thermal Boltzmann factor is along the positive real
  axis, and the quantum time evolution is along the unit circle. One
  of the characteristics of the complex dynamics constituted by
  repeated applications of RG is the Julia set, which determines the
  phase transitions.  We show that special boundary conditions can
  alter the nature of the transitions, and verify the claim for the
  one-dimensional system by transfer matrix calculations.  In two
  dimensions, there are alternating symmetry-breaking and restoring
  transitions, both of which are first-order, despite the criticality
  of the Curie point.  In addition, there are finer structures because
  of the fractal nature of the Julia set. Our approach can be extended
  to multi-variable problems, higher dimensions, and approximate RG
  transformations expressed as rational functions.
\end{abstract}

\maketitle

\section{Introduction}

A dynamical quantum phase transition (DQPT) is a sudden or
non-analytic change in the behaviour of a large quantum system during
its time evolution.  A typical example is the Loschmidt echo, which
measures the probability $P(t)$ of finding the system in the original
state $\psi_0$ after time $t$ \cite{heylreview,amina}.  Interestingly,
these transitions occur at critical times without having to  change any of
the system's parameters.

Quench dynamics in quantum systems is a widely researched topic that
holds importance in several areas.  It ranges from conceptual issues
such as dynamics in many-body systems, thermalization, entanglements,
and novel dynamics to practical applications in quantum computations
and information \cite{krish,paraj1,paraj2}.  While DQPT was originally
observed in the transverse field Ising model \cite{heylprl,heyl2},
subsequent investigations have been conducted on various types of
systems for both pure and mixed states
\cite{Andras,dmtn,ks17,sun2023,zvyagin,pozsgay,expt2,experiment}. Different
types of quenches have been studied in Floquet systems, topological
models, bosonic and fermionic systems, and many others
\cite{jafari1,jafari2,jafari3,maslowski,galitski,sun,chen,cao,ding,tian2020,halimeh,niccolo}.
A crucial question that arises is whether the phases, transitions, or
criticalities in DQPT are just analogous to the known thermal phases
and transitions observed in the same system or {\it new phases or
  transitions may emerge  in the quantum domain}.

The terminology used here follows thermodynamic definitions.  A system
is deemed to be in the same phase at two different times if its
properties evolve smoothly whereas a transition between phases happens
when the pattern of evolution experiences abrupt or divergent changes
in time derivatives \cite{commdiv}.  DQPT is classified as a
first-order transition if the first derivative of $P(t)$ (or rather,
$-\ln P(t)$) in time is discontinuous at the transition point.  Any
other form of singular behavior is regarded as a continuous
transition.

\subsection{On DQPT}
Consider the time evolution of a quantum system described by the
Hamiltonian $H$ when started in a state $|\psi_0\rangle$, which is not
an eigenstate of $H$.  The state at time $t$ is $|\psi_t\rangle=
e^{-iHt} |\psi_0\rangle$ (take $\hbar=1$).  The quantity of interest
is the probability of finding the system in the initial state, called
the Loschmidt echo, and it is given by
\begin{equation}
    \label{eq:1}
    P(t)=|L(t)|^2, \quad {\rm where}\quad L(t)=\langle\psi_0|\psi_t\rangle.
\end{equation}
In the basis of ${\cal W}$ orthonormal eigenstates of
$H$,
\begin{equation}
  H=\sum_{n=1}^{{\cal W}} E_n |n\rangle\langle{n}|, 
\end{equation}
with $\langle n|m\rangle=\delta_{nm},$
where $\delta_{nm}$  is the Kronecker-delta,
the states at time $t=0$ and $t$ are, respectively, 
\begin{subequations}
\begin{equation}
\label{eq:2}
|\psi_0\rangle=\sum_{n=1}^{{\cal W}} c_n |n\rangle, {\ \rm and\ }  |\psi_t\rangle = \sum_{n=1}^{{\cal W}} c_n e^{-iE_n t}|n\rangle,
\end{equation}
so that 
\begin{equation}
\label{eq:3}
L(t)=\sum_{n=1}^{\cal W} |c_n|^2 e^{-iE_n t},
\end{equation}
\end{subequations}
{{with $|c_n|^2$ as the probability of the $n$th state at $t=0$.}}

Let us restrict ourselves to the case with $c_n=1/{\sqrt{\cal W}}$,
i.e., all eigenstates are equally probable in the initial state. 
The Loschmidt amplitude is then
 $L(t)=\frac{1}{{\cal W}} \sum_{n=1}^{\cal W} e^{-iE_n t}$
which is to be compared with the partition function at the inverse
temperature $\beta$, $Z(\beta)=\sum_n e^{-\beta E_n}$.

We define a complex function 
\begin{subequations}
\begin{equation}
    \label{eq:4}
    {\cal L}(z)=\frac{1}{{\cal W}} \sum_{n=1}^{\cal W} e^{-z E_n },
\end{equation}
such that 
\begin{equation}
    \label{eq:5}
    {\cal L}(z)=\left \{ \begin{array}{ll} 
      \frac{1}{{\cal W}}  Z(\beta), &{\textrm{ for}} \ z=\beta {\textrm{ (real)}}, \\[6pt]
       L(t),  & {\textrm{ for }} z=i t {\textrm{ (real}}\  t).
    \end{array} \right.
\end{equation}
\end{subequations}
In simpler terms, ${\cal L}(z)$ is the extension of the thermal
partition function of a quantum system from the real axis to the
complex plane by extending the real inverse temperature $\beta$ to the
complex variable $z$.  This analytic continuation connects the
partition function to the Loschmidt amplitude.

The extensivity of thermodynamic quantities requires that the free
energy $\beta F=-\ln Z$ is proportional to the size of the system so
that for $N$ degrees of freedom, $N^{-1} \ln Z$ in the limit
$N\to\infty$ should be independent of $N$.  As an analytic
continuation, we then expect $N^{-1} \ln L$ to be also independent of
$N$ so that (Re is real part)
\begin{equation}
  \label{eq:70}
  P(t)=e^{-2N f},{\textrm{ where }} f=-{\textrm{ Re }}\lim_{N\to\infty} N^{-1} \ln L,
\end{equation}
where $f$ is independent of $N$, and is called the rate function in
the large-deviation theory \cite{largedev}. The phase transitions are
given by the singularities of $f$.

It is convenient to use $y=e^{\beta \Delta}$ (a Boltzmann factor with
$e^{it \Delta}$ as its complex extension) as the variable, instead of
$\beta$ and $t$, and treat $L, Z, {\cal L}$ as functions of $y$.
Here, $\Delta >0$ is a typical energy scale for the problem. In the
complex-$y$ plane, the thermal partition function is defined along the
positive real axis while the quantum time evolution is along the unit
circle $|y|=1$. Specifically, the singularities in $f$ as $y$ varies
on the unit circle are the DQPTs during the quantum evolution.

\subsection{Objectives} \label{sec:purpose}

We can use the mathematical approach of zeros of the partition
function $Z(y)$ in the complex-$y$ plane to generate the singularities
in free energy \cite{huang,mfisher}. It is important to note that, in
any finite system, there are no real positive zeros of $Z(y)$ (no
phase transition).  However, in the thermodynamic limit (infinite
size), the zeros may pinch the real axis as a limit point, creating a
thermal phase transition. Additionally, the same set of zeros also
describes ${\cal L}(y)$ (and, so, $L(y)$).  Therefore, the local
dispositions of the zeros (not just pinching at a point) around the
unit circle $|y|=1$ determine the singularities in $f(y)$ for the
Loschmidt echo, connecting the complex zeros to DQPT. A thermodynamic
limit is required to ensure a dense set of zeros, without which there
can be no phase transitions.

A different but more general framework for analyzing a many-body
interacting system is the renormalization group (RG) method
\cite{huang,smb-serc}, which is the method of choice for DQPT.  The technique
involves two essential steps. Firstly, small-scale degrees of freedom
(df) are integrated out, and their effects are accounted for by
adjusting (renormalizing) the effective interactions of the remaining
dfs.  Secondly, the remaining system is rescaled to keep physical
quantities invariant, like the free energy per df. This leads to the
transformation of parameters, where $y$ becomes $y'=R(y)$, and the
concerned physical quantity is renormalized.  By repeatedly performing
these steps, the large-scale behavior of the system can be determined
by observing how the renormalized parameters approach their fixed
point values. The stable fixed points represent the phases of the
system, while the unstable ones represent phase transitions or
critical points. Apart from these, the renormalization of the free
energy allows us to generate the zeros of the partition function in
the thermodynamic limit.

This paper explores the relationship between DQPT and the transitions
in the equilibrium thermal problem to uncover the possibilities of new
phases and transitions. The connection is via the RG transformation in
the complex plane and the complex zeros of the partition function in a
{\it class of models amenable to exact analysis}. We study the 3-state
Potts model \cite{wu,ks17,affleck,dai,chepiga} defined on
hierarchical lattices for which the real-space renormalization group
can be implemented exactly, thereby avoiding artifacts of
approximations \cite{hier,derrida,berker}.  The repeated RG
transformations constitute an {\it iterated map in the complex plane}
of a properly defined Boltzmann factor.  Results of the dynamics of
one complex variable \cite{milnor,beardon,carleson} (complex dynamics
in short) can be made to bear upon the quantum problem.  Near-exact
numerical computations then supplement these analytical results.

\subsection{Outline}\label{sec:outline} 
{{The 3-state quantum Potts model on a lattice
    is defined in Section \ref{sec:potts-model}, where 
    the eigenstates and eigenvalues required for the subsequent analysis can be
    found. The hierarchical lattices used in this paper and their
    characteristics  are in Section
    \ref{sec:hier-latt}.  The particular quench of interest here is
    elaborated in Sec. \ref{sec:quench} while the early time behaviour of
    the rate function for the Loschmidt echo after the quench is analyzed in Sec.
    \ref{sec:small-time-behaviour}.

The details of the RG procedure is discussed in
Sec. \ref{sec:impl-rg-compl}.  The  exact implementation of the RG
transformation allows us to compute the rate function as a sum along
the RG trajectory, given by Eq. (\ref{eq:26}). 
  In this process, a boundary-dependent term emerges
that plays a critical role in the one-dimensional case.

The methods of complex dynamics for iterated maps in the complex plane
are summarized in Sec. \ref{sec:complex-dynamics-one}.  The phases and
the phase transitions in the system are determined by the flows
in the complex plane, which 
fall into two categories: the Fatou set, controlled by attractive fixed points
and the Julia set forming the boundary of the Fatou
set. The properties of these sets are summarized in Sec.
\ref{sec:complex-dynamics-one}.  The Julia set is linked to the zeros
of the partition function and the rate function in Sec.
\ref{sec:zeros-julia-sets}.

DQPT in the one-dimensional Potts model is discussed in detail,
particularly the role of the Julia set, in Sec.
\ref{sec:one-dimens-potts}.  We also examine the effects of boundary
conditions (Sec. \ref{sec:open-bound-cond}-D) and further corroborate
them using a transfer matrix approach (Sec.
\ref{sec:interm-phase-transf}).  In Section \ref{sec:quantum-case}, we
provide a statement on the conditions under which boundaries may
supress an otherwise allowed DQPT.  Moving on to the two-dimensional
case in Section \ref{sec:higher-dimensions-d1}, we observe that,
despite the complications introduced by the wide intersections of the
unit circle with the Julia set, DQPTs are first-order and described by
the repulsive fixed points of the Julia set.  The paper concludes in
Sec. \ref{sec:conclusion} with a summary and remarks on possible
generalizations of the approach developed in the paper. 

A few algebraic details can be found in the Appendices. A few 
periodic orbits for the one-dimensional Potts chain with their
multiplicities are listed in 
Appendix \ref{sec:peri-orbits-potts}. The full Julia set for the
two-dimensional case is shown in Appendix \ref{sec:julia-set-b=2}.  }}

\section{The Potts model}\label{sec:potts-model}   
We study the three-state ($q=3$) Potts model on a lattice (Fig.
\ref{fig:latt}) with nearest-neighbour
interaction \cite{wu,affleck,dai,chepiga}.  In the quantum
model, each site has a Potts variable with eigenstates represented by
a planar spin oriented in three symmetric directions: $0, 2\pi/3,
4\pi/3$ on a circle (Fig. \ref{fig:circle}c). The ferromagnetic
interaction generates a three-fold degenerate ground state, which may
endure quantum and thermal fluctuations to produce an ordered state.
As a result, the model may demonstrate a symmetry-breaking transition
(breaking of the discrete permutation symmetry in this case).

The Hamiltonian ${\cal H}$ has two parts, (i) an interaction term $H$
that tries to order the spins, and (ii) a `transverse-field' term
$H_{\Gamma}$ that disrupts ordering by flipping the spins, and these
are given by \cite{affleck}
\begin{subequations}
\begin{eqnarray}
    {\cal H}&=&H+H_{\Gamma},\label{eq:7}\\
    H&=&-\frac{{\cal J}}2 \sum_{<jk>}( {\bm{\Omega}}_j^{\dagger}
    {\bm{\Omega}}_k+ {\bm{\Omega}}_k^{\dagger} {\bm{\Omega}}_j),\quad
    ({\cal J}>0), \label{eq:8}\\
    H_{\Gamma}&=&-\Gamma \sum_j {\bm{T}}_j,\label{eq:9}
\end{eqnarray}
where  dagger denotes Hermitian conjugate, 
$j,k$ denote the lattice sites with $<\cdots>$ denoting
nearest-neighbours,
\begin{eqnarray}
    {\bm{\Omega}}_j&=& {\bm{I}}\otimes {\bm{I}} \otimes
    \cdots{\overbrace{\bm{\Omega}}^{j{\rm th}}}\otimes {\bm{I}}\otimes\cdots,\label{eq:10} \\
    {\bm{T}}_j&=& {\bm{I}}\otimes {\bm{I}} \otimes
    \cdots{\underbrace{\bm{T}}_{j{\rm th}}}\otimes {\bm{I}}\otimes\cdots,\label{eq:11}
\end{eqnarray}
\end{subequations}
with ${\bm{\Omega, T}}$ at the $j$th position in the direct products
over sites, and ${\bm{I}}$ the identity. The Potts variables are given
by ($\omega=e^{i 2\pi/3}$)
\begin{equation}
\label{eq:6}
    {\mathbf{\Omega}}=\left(\begin{array}{ccc}
    1&0&0\\
    0&\omega&0\\
    0&0&\omega^{2}
    \end{array}\right),\, {\textrm{ and }}
  {\mathbf{M}}=\left(\begin{array}{ccc}
    0&1&0\\
    0&0&1\\
    1&0&0
    \end{array}\right),
\end{equation}
with $ {\bm{T=M+M^{\dagger}}}.$ Here, ${\bm{\Omega, T}}$ for the Potts
spins are expressed in the eigenstates $|1\rangle,$ $|2\rangle,$ $|3
\rangle$, and, in that basis, ${\bm{M}}$ is a spin flip operator,
satisfying
\begin{equation}
  \label{eq:69}
{\bm{\Omega}}^q={\bm{M}}^q={\bm{I}}, {\textrm{ with }} q=3.
\end{equation}

\hlattice

The eigenstate for the largest eigenvalue of ${\bm{T}}$ is
$$|0\rangle=\frac{1}{\sqrt{3}} (|1\rangle +|2\rangle+|3\rangle).$$  
The special property of ${\bm{T}}$ is that it flips any of the three
states $|\mu\rangle, \mu=1, 2, 3$ to an equal superposition of the
other two states, e.g.,
\begin{equation}
  {\bm{T}} |1\rangle= |2\rangle+|3\rangle, 
\label{eq:12} 
\end{equation}

The $3^N$ eigenstates of $H$ for $N$ spins, $H\,|n\rangle=E_n
\,|n\rangle,$ are the direct product states of individual spin states
$|\alpha\rangle, \alpha=1,2,3$ as
\begin{subequations}
  \begin{equation}
    \label{eq:13}
    |n\rangle=\otimes_{j}|\alpha\rangle_j \equiv |\alpha_1\alpha_2\cdots\alpha_N\rangle,
  \end{equation}
with eigen-values $E_n$ given by 
\begin{equation}
   E_n= -{\cal J}\sum_{<jk>} \cos(\theta_j-\theta_k),
\qquad\theta_j=\frac{2\pi}{3} (\alpha_j-1).
\end{equation}
\end{subequations}

The ground state for a single pair of spins is a state with parallel
spins, $\theta_j=\theta_k$, and has energy $-{\cal J}$.  This state is
threefold degenerate. On the other hand, the excited state is six-fold
degenerate and is achieved through nonparallel spins with
$|\theta_j-\theta_k|=2\pi/3 \mod 2\pi$, with an energy of $-{\cal
  J}/2$.  To simplify our calculation, we shift and rescale the energy
to express it as
\begin{equation}
  \label{eq:14}
  E_n= -J \sum_{<jk>} \delta_{\alpha_j,\alpha_k},
\end{equation}
so that the ground state energy is $-J$ and the gap in the spectrum
for a bond is $J$. The Boltzmann factor
\begin{equation}
  \label{eq:77}
  y=e^{\beta J},
\end{equation}
appears as the variable of choice.

\subsection{Hierarchical lattices}\label{sec:hier-latt}
The Potts spins are placed on the sites of a lattice constructed in a
manifestly scale-invariant way, as shown in Fig. \ref{fig:latt}. The
nearest neighbours are defined by the bonds.  We start with a single
bond with two sites as generation $n=0$ and a diamond-like motif of
$2b$ branches (Fig. \ref{fig:latt}b) \cite{diam}.  The lattice for
generation $n$ can be constructed from the $(n-1)$-generation lattice
by replacing each bond with the diamond motif.  When $b=1$, the
recursive process results in a one-dimensional lattice (Fig.
\ref{fig:latt}a).  From the growth of the lattice (the number of bonds
$B$ with generation $n$), the Hausdorff dimension \cite{dimen} of the
lattice is $d_b=\frac{\ln 2b}{\ln 2}$. This expression gives $d_1=1$,
which is consistent with Fig. \ref{fig:latt}a, and $d_2=2$ for $b=2$.
We consider these two cases, viz., one-dimensional ($b=1$) and the
two-dimensional ($b=2$) lattices.

The numbers of bonds and sites for generation $n$ are
\begin{equation}
  \label{eq:73}
   B_n=(2b)^{n},{\textrm{ and }}  N_n=2+ b \frac{(2b)^n-1}{2b-1}, 
\end{equation}
with $\lim_{n\to\infty} \frac{N_n}{B_n}=\frac{b}{2b-1}.$ Thanks to
this asymptotic proportionality, we use the number of bonds (instead
of the number of spins) to describe extensivity.

The recursive construction of the lattice allows one to implement
real-space RG exactly on these lattices \cite{hier,derrida}.
Moreover, dimensionality can be tuned by adjusting the value of $b$,
which allows for a comprehensive examination of the impact of
dimensionality on relevant properties. Recent studies have
predominantly focused on quantum phase transitions in higher
dimensional ($>1$) Euclidean lattices for the $q$-state quantum Potts
model \cite{dai,chepiga}. Although mapping the $d$-dimensional quantum
problem to a $(d+1)$-dimensional classical statistical mechanical
problem has proven advantageous, such mappings are of limited use for
the hierarchical lattices.

\subsection{The quench and the echo}
\label{sec:quench}

We are interested in the sudden quench from $\Gamma\to\infty$ to $\Gamma=0$.  
A large field, $\Gamma\to\infty$,
forces the spins to be in the eigenstate of ${\bm{T}}$ with the
largest eigenvalue.  This pure  state is in a product form 
\begin{equation}
  \label{eq:15}
  |\psi_0\rangle=\bigotimes_{j}\ |0\rangle_j=\bigotimes_{j}\ \left[\frac{1}{\sqrt{3}} (|1\rangle +|2\rangle+|3\rangle)\right]_j.
\end{equation}
At time $t=0$, $\Gamma$ is switched off, and the Potts system
undergoes a unitary evolution with $H$, Eq. (\ref{eq:8}).  Our aim is
to determine the behaviour of ${\cal L}(y)$ (Eq. (\ref{eq:4}) and
$f(y)$ defined in Eq. (\ref{eq:70}), by extending the Boltzmann factor
to a complex number
\begin{equation}
  \label{eq:19}
  y=\exp(z J), {\text{ \ with \ }} zJ= \beta J, {\textrm{ or }}  i t J/\hbar,
\end{equation}
where all the dimensionwala quantities are shown explicitly. In the
complex-$y$ plane, $y=1$ corresponds to the infinite temperature case
and the $t=0$ state. Henceforth, we set $\hbar=k_B=1$

\section{Early-time behaviour }
\label{sec:small-time-behaviour}

The normalization conditions ensure that $f(t=0)=0$.  In the
complex-$y$ plane, $f$ is analytic around $y=1$ which corresponds to
$t=\beta=0.$ For small values of $t$, a Taylor series expansion of
$f(t)$ can be performed, and analyticity around $y=1$ guarantees that
the derivatives there are independent of the direction in the complex
plane (Cauchy-Riemann conditions).  As a result, the thermal
free-energy's high-temperature behavior determines the behavior of
$f(t)$ at early times.

Expanding the exponentials in Eq. (\ref{eq:3}), 
$$e^{iE_n t}=1 + i E_n t - \frac{1}{2} E_n^2 t^2+....,$$
we can write the rate function or the free energy (Eq. (\ref{eq:70})
per bond as
\begin{subequations}
\begin{equation}
  \label{eq:44}
  f(t)= \frac{\langle E^2\rangle -\langle E\rangle^2}{B}\, t^2 + O(t^3),
\end{equation}
where the averaging $\langle\cdots\rangle$ is done over the
probability distribution at $t=0$ {{(see Eq. \ref{eq:3})}}. 
There is no linear term in $t$
because $f$ involves only the real part. The quadratic dependence on
$t$ is consistent with time reversibility under the unitary evolution
of the initial state with $H$.

The specific heat (per bond) is given by the energy fluctuation
\begin{equation}
  \label{eq:45}
  c(\beta)= \frac{\langle E^2\rangle_{\beta} -\langle
  E\rangle_{\beta}^2}{B}  \beta^2,
\end{equation}
where $\langle\cdots\rangle_{\beta}$ denotes the thermal average at
inverse temp $\beta$.  For $\beta\to 0$, $c(\beta)\sim \beta^2$
because the energy spectrum for the model is bounded, and the variance
is extensive (i.e., $\propto B$).  In other words, if the high
temperature specific heat
$$c\approx \frac{C_0}{T^2},{\textrm{  such that }} C_0=\lim_{T\to\infty}
T^2 c(T),$$
the initial $t$ dependence of the rate function during the quantum quench is
\begin{equation}
  \label{eq:46}  f(t)\approx C_0 t^2, {\textrm{  (small }}  t), C_0>0.
\end{equation}
\end{subequations}
Eq. (\ref{eq:46}) connecting the quench behaviour to the specific heat
is a general result valid for any $H$ with bounded spectrum, provided
the quench is from a state of uniform distribution of states.

\section{Implementing RG in the complex  plane}\label{sec:impl-rg-compl}

{{The procedure adopted in this paper is the real-space
    renormalization group.  The scale invariance of the hierarchical
    lattices (Sec.  \ref{sec:hier-latt}) allows us to decimate
    (integrate out a set of degrees of freedom) exactly and
    repeatedly. The exact decimation gives us the renormalized
    parameters for the decimated lattice, as well as the
    transformation of the rate function.

Below is a summary of the basic steps, which are further illustrated
through diagrams in Appendix \ref{sec:rg-equations}. The rate function
is also expressed as a sum over the renormalized parameters (here
$y$), with an additional boundary term that is needed in our
discussion of the one-dimensional case.  }}

The RG analysis involves a partial summation of a subset of spins to
define a new effective parameter for the remaining spins.  The
protocol would be as follows \cite{hier,derrida}.
\begin{enumerate}
\item Let us start with a large lattice and sum over the spins
  represented by the red squares in Fig. \ref{fig:latt}.

\item The partial partition functions correspond to the weights
  (Boltzmann factors for the thermal problem) for the remaining spins.
  In our case, the $b$-branched diamond reduces to a single bond
  connecting the blue spins. (The {\em decimation step})

\item The partial partition function of a diamond of $b$ branches with
  the two blue spins in the same state, say state $\alpha=1$ is
  $Z_{\rm dia}(y|\alpha,\alpha)=(y^2+2)^b$ which is proportional to
  the partition function of a bond with the renormalized weight $y'$,
  $Z_{\rm bond}(y'|1,1)=y'$. There are $q$ such partition functions,
  and all are equal.

\item The partial partition function for the blue spins in different
  states, $\alpha$ and $\beta\neq\alpha$ is $Z_{\rm
    dia}(y|\alpha,\beta)= (2y+1)^b$ which should be proportional to
  the weight of a renormalized bond connecting two different spins,
  $Z_{\rm bond}(y'|\alpha,\beta)=1$. By symmetry, there are $q(q-1)$
  such partition functions, and all are equal.

\item There are two RG equations to determine the renormalized
  parameter $y'$ and the proportionality factor $c(y)$.  The RG
  equations are
\begin{subequations}
  \begin{equation}
    \label{eq:20}
    (y^{2}+2)^b=cy', {\text{ and }} (2y+1)^b=c,
  \end{equation}
and, therefore,
\begin{equation}
  \label{eq:21}
  y'=R(y)=\left(\frac{y^2+2}{2y+1}\right)^b,  {\text{ and }} c= (2y+1)^b
\end{equation}
\end{subequations}
This completes the {\em  rescaling step}.

\item The total partition function, with $q=3$ is  
\begin{subequations}
  \begin{equation}
    \label{eq:22}
    Z(y)=q Z(y|\alpha,\alpha)+q(q-1) Z(y|\alpha,\beta),
  \end{equation}
  which should be invariant under the RG transformation.  The
  invariance requirement is satisfied by the proportionality factor
  $c(y)$.  The partition function for the $n$th generation $Z_n(y)$ is
  related to $Z_{n-1}(y')$ by
 \begin{flalign}
   \label{eq:23}
   Z_n(y)&= Z_{n-1}(y')\, c(y)^{B_{n-1}},\\
  \intertext{ with\  }
   Z_1(y)&= q(y+q-1)\label{eq:24}
 \end{flalign}
\end{subequations}

\item The Loschmidt rate function per bond is calculated along the RG
  trajectory from the repeated application of $R(y)$.  We denote the
  $n$th iteration by \cite{commfo}
  \begin{subequations}
 \begin{flalign}
    R^{(n)}(y)&\equiv R(R(...R(y)))),\label{eq:79}\\
   \intertext{and use \ }
 y^{(j)}&\equiv R^{(j)}(y), {\textrm{ with }}
   y^{(1)}=y', y^{(0)}=y. \label{eq:80}
 \end{flalign}
\end{subequations}
 The rate function is then  given by (in the full complex form)
\begin{subequations}
\label{eq:40}
  \begin{align}
    f_n(y)&= -\frac{1}{B_n} \ln Z_n(y)\label{eq:78}\\
          &= \frac{1}{2b}
               f_{n-1}(y')+\frac{1}{2b} g(y), \quad (g(y)=\ln c(y)),    \label{eq:25}\\
          &=  \sum_{j=0}\frac{1}{(2b)^j} g(y^{(j)})
          +\frac{1}{(2b)^n} \ln Z_1(y^{(n)}). \label{eq:26}
  \end{align}
\end{subequations}
\end{enumerate}
Eq. (\ref{eq:26}) is rapidly convergent and can be used to compute the
Loschmidt echo numerically for large $n$.  We obtain $f(y)$ by taking
the real part at the end.

The principal branch of the log function is to be used (cut along the
negative real axis).  However, as the Loschmidt echo is determined by
the real part of $f(y)$ (Eq.  (\ref{eq:40}), and the multivaluedness
of the log function appears in the imaginary part only, the location
of the branch cut is not crucial for the problem at hand.

\section{Complex dynamics in one variable}\label{sec:complex-dynamics-one}
The gradual thinning of degrees of freedom in the renormalization
group approach, as discussed in Sec. \ref{sec:impl-rg-compl}, requires
repeated applications of the transformation $y'=R(y)$ of the
characteristic Boltzmann factor.  To address the Loschmidt echo
problem, we treat $y$ as a complex variable in the extended complex
plane ${\widehat{\cal C}}={\cal C}\cup \{\infty\}$ which is the
Riemann sphere obtained via stereographic projection \cite{commdist}.
Infinity has to be included because the zero temperature corresponds
to $\beta, y \to\infty$.

{{In contrast to the thermal problem (real $y$), where fixed points
    are sufficient, the RG flows in the complex plane require a more
    detailed understanding of their behavior. These flows can be
    divided into two categories: the Fatou set and the Julia set. The
    Fatou set represents the set of points that converge to one of the
    transformation's attractive fixed points. This set can be
    considered as the potential phases of the system, but only if the
    attractive fixed points are accessible from the unit circle
    ($|y|=1$). On the other hand, the Julia set is composed of the
    boundaries of the basins of attractions and includes the repulsive
    fixed points. It generally contains an infinite number of points
    and is a dense set. }}

{{The study of repeated maps in the complex plane is made possible
    by the use of complex dynamics methods.  This section discusses
    important aspects such as the classification of fixed points and
    orbits (Sec. \ref{sec:fixed-points}), the characteristics of the
    Julia and the Fatou sets (Sec. \ref{sec:julia-set-fatou}),
    providing an overview of the methods and results of complex
    dynamics.  }}

\subsection{RG as an iterated map}\label{sec:rg-as-an}
The exact RG transformation, Eq. (\ref{eq:21}), is a rational function
$R={\cal P/Q}$, where ${\cal P}$ and ${\cal Q}$ are polynomials in
$y$.  This is because the RG transformation involves partition
functions of a small or finite number of degrees of freedom, and is,
therefore, nonsingular for real $y$.  In the complex-$y$ plane, there
can, however, be poles.  The iterated maps of such a function are
generally called dynamics in one complex variable, in short, {\em
  complex dynamics}.  A few general properties of such complex
dynamics are summarized here \cite{milnor,beardon,carleson}.

The rational form ${\cal P/Q}$ is such that there are no common
factors of ${\cal P}$ and ${\cal Q}$.  In this situation, if ${\cal
  P}$ and ${\cal Q}$ are polynomials of degree $p$ and $q$,
\begin{subequations}
\begin{equation}
  \label{eq:74}
  {\cal P}(y)=\sum_{j=0}^{p}a_j y^j, {\textrm{ and }} {\cal Q}(y)=\sum_{j=0}^{q}b_j y^j,
  (a_p,b_q\neq 0),
\end{equation}
the degree of the transformation $R$ is defined as 
\begin{equation}
  \label{eq:75}
{\textrm{deg}}(R)=\max(p,q),  
\end{equation}
\end{subequations}
so that there are ${\textrm{deg}}(R)+1$ fixed points (fp) in
${\widehat{\cal C}}$ as solutions of $R(y)=y,$ or $P(y)=y Q(y)$
(counting multiplicities and infinity).

For the thermal problem, with $y$ as a Boltzmann factor, $1\leq y\leq
\infty,$ there should be at least two fixed points, (i) $y=1$
representing the infinite-temperature fixed point (noninteracting
limit), and (ii) $y=\infty$ representing the zero-temperature
behaviour (generally an ordered state).  In addition, there could be a
critical fp representing a transition. Now, for the zero-temperature
fixed point, $y'=R(y)\to\infty$ as $y\to\infty,$ which implies $p>q.$
Consequently, the RG transformation should be such that
\begin{equation}
  \label{eq:76}
 p\geq 2, \quad p>q, {\textrm{ and }} {\textrm{deg}}(R)=p. 
\end{equation}

\subsection{Fixed points}
\label{sec:fixed-points}

Let us represent the fps by $y^*_k, k=1,2,...,{\textrm{deg}}(R)+1$.
Each fp $y^*_k$ is characterized by its multiplier (also called
eigenvalue)
\begin{subequations}
\begin{equation}
  \label{eq:32}
  \lambda=\left. \frac{dR(y)}{dy\ \ \,}\right|_{y^*_k},
\end{equation}
which can be used for a classification of the fps as
\begin{enumerate}
     \item Repelling (unstable) fp, if $|\lambda|>1,$
     \item Attracting (stable) fp, if $0<|\lambda|<1,$
     \item Super-attracting fp, if $\lambda=0$, i.e. $R(y)=y^*_k+ c_m
       (y-y^*_k)^m+...$, for some integer $m\geq 2$
     \item Neutral, if $\lambda=e^{i2\pi\theta}, |\lambda|=1,$ with
       three possibilities,
     \begin{enumerate}
         \item $\theta=0$, so that $R(y)=y+c_2 (y-y^*_k)^2+...,$
         \item  $\theta=m/n$ so that $\lambda^n=1$ for integers $m,n,$
         \item $\theta$ is not a rational fraction.
     \end{enumerate}
\end{enumerate}
Neutral points do not occur in this paper.

For an fp at $\infty, R(\infty)=\infty$, the multiplier is determined
by the behaviour of $\widetilde{y}=1/y$ with the transformed map
$1/R(1/\widetilde{y})$. The multiplier is then given by
\begin{equation}
  \label{eq:35}
  \lambda=\frac{1}{\left. \frac{dR(y)}{dy\ \ \,}\right|_{\infty}},
\end{equation}
which can be used for its classification in the above scheme. 
\subsection{Periodic orbits}
\label{sec:periodic-orbits}

The RG trajectory or orbit for any point $y_0$ is the sequence of
numbers, $y^{(n)}\equiv R^{(n)}(y_0)$ (see Eq. (\ref{eq:79})),
 \begin{equation}
  y_0 \underset{R}{\to}   y^{(1)}  \underset{R}\to y^{(2)} \cdots  \underset{R}\to y^{(n)} \cdots .  \label{eq:seq}
 \end{equation}
 If $y^{(n)}=y_0$ for some integer $n$, then the trajectory is
 periodic, a cycle of order $n$ for the smallest integer $n$.  The
 multiplier $\lambda$ for a cycle is given by
 \begin{equation}
   \label{eq:36}
 \lambda=\left .\frac{dR^{(n)}}{dy\quad}\right|_{y_0}
 =\prod_{j=0}^{j=n-1}\left. \frac{dR}{dy}\right|_{y^{(j)}},  
 \end{equation}
\end{subequations}
and the product formula follows from the chain rule. Note that
$\lambda$ is independent of the starting point $y_0$. The cycles can
then be classified by the value of the multiplier at the fixed points
of $R^{(n)}(y)$.  A periodic orbit (also called a cycle) indicates that
the RG transformation simplifies if it is based on a larger cell.
However, the small-scale periodicities would show some nice properties
at a local level.  We shall see such examples in this paper.

\subsection{Julia set and Fatou set}
\label{sec:julia-set-fatou}

The extended complex plane $\widehat{\mathbf{C}}$ (i.e., the Riemann
sphere) can be divided into two broad categories based on how the RG
trajectories approach the fps. One set, to be called the Fatou set,
consists of points whose RG trajectories approach any attractive fp or
periodic cycle.  This set includes the basins of attraction (not
necessarily connected) of all the attractive fps.  The second
category, named the Julia set, $J(R)$ of the transformation $R$, is
the boundary of the Fatou set, and $J(R)$ consists of points whose
trajectories may behave chaotically.  The trajectories remain confined
to the set.  The importance of the Julia set is discussed below, but
before that, we summarize \cite{milnor,beardon,carleson} some of the
properties of the two sets here.

\begin{enumerate}
\item Trajectories on ${\widehat{\mathbf C}}$ are considered
  equicontinuous when two points maintain their proximity throughout
  their paths \cite{equicont}.  All such equicontinuous trajectories
  belong to the Fatou set, which, as already mentioned, need not be a
  connected set. It should be noted, however, that any attractive
  fixed point possesses a connected neighbourhood, no matter how small,
  where the multiplier $\lambda$ can accurately depict the
  trajectories.

\item If the trajectories behave chaotically, i.e., there is a
  sensitive dependence on the initial point, and the equicontinuous
  definition \cite{equicont} fails, then $y_0$, the starting point,
  belongs to the Julia set. The points of the Julia set transform
  among themselves under $R$ and do not flow to any attracting fp.

\item The Fatou set is open, and the Julia set is its boundary. 
  The Julia set is dense  with dimension ${\rm dim} J\leq
  2$ as $J\subset {\widehat{\mathbf{C}}} $
\item All attracting fixed points and cycles belong to the Fatou set.
\item All repelling fps and periodic orbits belong to the Julia
  set.
\item If $y$ belongs to $J(R)$ then $R(y)\in J(R)$. It follows that
  $J(R^{(n)})=J(R)$, for any $n$. In other words, the Julia set of the
  $n$-fold iterate is the same as the Julia set of $R$.

\item If degree ${\textrm{deg}}(R) \geq 2$ (Eq. (\ref{eq:76})), then
  there are no isolated points in $J(R).$
\end{enumerate}

The Julia set is crucial in determining the phase transitions
discussed in this paper. Specifically, the unstable fixed points and
periodic points of the RG transformation, located at the intersections
of the Julia set with the unit circle or the positive real axis, are
of utmost importance.  Moreover, the stable fps belonging to the Fatou
set determine the phases of the system, provided that an RG trajectory
takes an initial point on the unit circle to that fp. Finally, the
Julia set is also connected to the zeros of the partition function.
This connection makes the intersections of the unit circle $|y|=1$
with the Julia set at points other than fixed points important as
possible candidates for singularities.

\section{Zeros and Julia sets}\label{sec:zeros-julia-sets}

The partition function for a finite lattice of $B_n$ bonds is a
polynomial in $y$ of degree $B_n$. The highest power of $y$ comes from
the threefold degenerate ground state, and this term is $3y^{B_n}$.
There are $B_n$ zeros of $Z_n$ in the complex-$y$ plane. We establish
the connection between the zeros of the partition function and the
Julia set through the renormalization group transformation
\cite{derrida}.

The partition function for the $n$th generation, with $B_n$ bonds, is
connected to that of the $(n-1)$th generation by the RG transformation
as given by Eq. (\ref{eq:23}). By denoting the zeros of $Z_n$ by
$\zeta_j$ and those of $Z_{n-1}$ by $\zeta'_k$, we rewrite Eq.
(\ref{eq:23}) as
\begin{subequations}
\label{eq:37}
\begin{align} 
  \prod_{j=1}^{B_n} (y-\zeta_j)& =  \left[\prod_{k=1}^{B_{n-1}}
    (y'-\zeta'_k)\right] c(y)^{B_{n-1}}, \label{eq:33} \\
              & = \prod_{k=1}^{B_{n-1}}
    \left[(y^2+2)^b-(2y+1)^b\zeta'_k\right],\label{eq:34}
\end{align}
\end{subequations}
where the second line follows from the substitution of $c(y)$ and $y'$
from Eqs. (\ref{eq:21}). Each factor of the right-hand side of Eq.
(\ref{eq:34}) is a polynomial of degree $2b$, which means that for a
given zero $\zeta'_k$ of $Z_{n-1}$, there are $2b$ roots of the
polynomial, which, in turn, represent the zeros of $Z_n$.  As a
result, $\zeta_j$, the solutions of $Z_n(y)=0$ are the roots of
$R(y)=\zeta'_k$ for every $\zeta'_k$, the solutions of $Z_{n-1}(y)=0$.
In short, $\zeta_j$ are the preimages of $\zeta'_k$.

By starting with the zeros of the smallest lattice, successive
preimages can be constructed.  In the limit of a large lattice (= a
large number of iterations), the preimages inevitably converge on the
Julia set of the map. This convergence on the Julia set is because the
preimages cannot be attractive fixed points.  The exceptional cases
are those where the initial zero is a fixed point of the RG
transformation. Such exceptional cases appear in the one-dimensional
case discussed below.
\section{One dimensional Potts model ($b=1$) }\label{sec:one-dimens-potts}

We now implement the RG procedure for the one-dimensional problem,
which corresponds to the $b=1$ case as shown in Fig. \ref{fig:latt}.
Two cases are to be distinguished, (i) an open chain, constructed as
in Fig. \ref{fig:latt}, and (ii) a periodic chain (pbc: periodic
boundary condition). 

In the following text, we discuss the RG transformation and its fp
behavior. We then use this transformation to examine chains with open
boundary conditions (Sec. \ref{sec:open-bound-cond}) and periodic
boundary conditions (Sec. \ref{sec:peri-boundry-cond}).  Although the
RG flow remains the same, the DQPT behavior differs, and we explain
this aspect in Sec. \ref{sec:contr-with-rg}. In Sec.
\ref{sec:open-versus-periodic-1}, we elaborate on the differences
between the thermal case and the quantum case, especially the role of
boundary conditions. Furthermore, we explore the nature of DQPT in
Sec. \ref{sec:nature-transitions}.  In Sec.
\ref{sec:interm-phase-transf}, we utilize a transfer matrix approach
\cite{huang} to support and augment the results from complex dynamics.

The RG transformation $y'=R(y)$  is given by a rational function (see
Eq. (\ref{eq:21}) and $b=1$)
\begin{equation}
  \label{eq:29}
  R(y)= \frac{y^2+2}{2y+1},
\end{equation}
which is of degree ${\textrm{deg}}(R)=2$ (see Eq. (\ref{eq:75})).
There are, therefore, only three fixed points, which are given below
with their multipliers ($\lambda$), as
\begin{equation}
  \label{eq:30}
 y^*=\left\{ \begin{array}{rll}
                 1, &(\lambda=0),& {\textrm{the high  temperature fp}}\\
                \infty, &(\lambda=2),& {\textrm{the zero temperature fp }}   \\
                -2, &(\lambda=0),& {\textrm{unphysical fp }}
                \end{array} 
               \right.
\end{equation}
See Fig. \ref{fig:onedcircle}.

\figboneqthree 

For the thermal behaviour, $y^*=1$ is the high temperature fixed point
that corresponds to the disordered state, while $y^*=\infty$
corresponds to the zero temperature three-fold degenerate ordered
state.  The third f.p., though it plays no role in the thermal
behaviour, is important for the quantum problem. Incidentally, the
magnitude of this fp ($|y^*|=2$) represents the multiplicity of the
spin states that do not contribute to the energy \cite{qm2}.

In the RG framework, the stable fixed points ($\lambda<1$) represent
the phases of the system, while the unstable points are the points of
phase transitions (thermodynamic critical points). The stabilities are
determined by the multipliers of the fps. The zero temperature ordered
state is unstable ($\lambda >1$) in one dimension, consistent with the
Landau argument \cite{huang} for any system with discrete symmetry.
The flow along the real axis to the attracting fp at $y^*=1$ indicates
the lack of any phase transition, or, equivalently, the lack of any
ordered state at any non-zero temperature.

In the complex-$y$ plane, $R(y)$ has a pole at $y_0=-\frac{1}{2}$, such that
$R(y_0)=\infty$. Moreover, for any $y_{\eta}=-\frac{1}{2} + i \eta,$
(real $\eta$)
\begin{equation}
  \label{eq:31}
 R(y_{\eta})=-\frac{1}{2} +i\eta', \textrm{ where } \eta'=R_p(\eta)=-\frac{9}{8\eta} +\frac{\eta}{2}. 
\end{equation}
As the RG trajectory of any point on this line stays on the line, the
map can also be described by $R_p$ for real arguments.  One notes that
the two imaginary fixed points ($\eta^*=\pm i 3/2$) of $R_p(\eta)$ are
just the two finite fps ($y^*=-\frac{1}{2}+i\eta^*=1,-2$) of Eq.
(\ref{eq:30}).

\subsection{Open boundary condition}
\label{sec:open-bound-cond}

For the open boundary condition, the partition function for a single
bond, Eq. (\ref{eq:24}) has a zero at $y=-2,$ which also happens to be
a fp.  Therefore, the partition function for the open chain, as
discussed in Sec. \ref{sec:zeros-julia-sets}, has a multiple zero at
the isolated point $y=-2$.  The Loschmidt amplitude for $B$ bonds, by Eq.
(\ref{eq:37}), is given by
\begin{subequations}
\begin{equation}
  \label{eq:38}
  L(t)=\frac{1}{3^B} (y+2)^B,
\end{equation}
with the rate function, Eq.(\ref{eq:40}), for $y=e^{iJt}$  given by
\begin{equation}
  \label{eq:39}
  f(t)= - {\textrm{ Re }} \ln \frac{y+2}{3}=- \frac{1}{2}\ln \frac{5+ 4 \cos Jt}{9}.
\end{equation}
\end{subequations}
See Fig. \ref{fig:b1ft} for a plot of $f(t).$ There is no DQPT in an
open chain, as was shown in Ref \cite{amina}.  The duality
transformation of the one-dimensional classical chain where each bond
acts as a Potts spin in a field with no interaction, leads to an
innocuous oscillation characterized by one isolated zero of the
partition function.

\figboneopen

\subsubsection{Contradiction with RG?}
\label{sec:contr-with-rg}

The absence of any singularity in $f(t)$ apparently contradicts the
results of RG, which, from the flow to the fixed points, predict phase
transitions at $A_1$ and $A_2$. The clue is in the boundary term in
Eq. (\ref{eq:26}), which is determined by the partition function of the
smallest structure at the renormalized $y$.  We show that the boundary
term itself is singular, and there is a perfect cancellation of the
singularity from the sum on the rhs of Eq.  (\ref{eq:26}).  More
detailed discussion is given in Sec.  \ref{sec:importance-freedom}.

\perodlat

\subsection{Periodic boundary condition}
\label{sec:peri-boundry-cond}

For the periodic boundary condition, the construction starts with the
triangle, Fig. \ref{fig:pbc}, whose partition function is given by
\begin{equation}
  \label{eq:41}
  Z_{\Delta}=3 y^3+18 y + 6,
\end{equation}
with a total of $27$ states (i.e., $Z_{\Delta}=27$ for $y=1$). The preimages of
the three zeros,
\begin{equation}
  \label{eq:42}
-0.32748000207..., 0.16374000103...\pm i 2.4658532729...,
\end{equation}
on successive back iterations, approach either infinity or a point on
the line $z=-1/2 + i \eta$ (Eq. (\ref{eq:31})), which is the Julia set
for $R(y)$.  As the point at infinity (a single point on the Riemann
sphere) is unstable (repulsive), it belongs to the Julia set as well.

The points on the complex plane are separated into two groups by the
Julia set. These groups belong to either the basin of attraction for
the attracting fixed point $y^*=1$, or the basin for the attracting
point $y^*=-2$. Incidentally, these are the only two critical points
of the map.

The Julia set intersects the unit circle at two points, namely
$A_1=e^{i2\pi/3}, A_2=e^{i4\pi/3}$ (Fig. \ref{fig:onedcircle}).  These
two points form a {\em periodic cycle of order two}. There is a
uniform density of zeros along the vertical line at the intersection
point.  Such a uniform density is a signature of a first-order
transition which is indicated by a slope discontinuity of the rate
function.

\figonedlosch

We now combine the results to see the behaviour of the Loschmidt echo.
At time $t=0$, the rate function $f(t)=0$.  As time progresses, the
system moves towards $A_1$ along the unit circle in the anticlockwise
direction, Fig.  \ref{fig:onedcircle}, but the rate function is just
an analytic continuation of the initial state described by the
attracting fp $y^*=1$ (high temperature, disordered state).  Except
for the time interval from A$_1$ to A$_2$, i.e. between
\begin{equation}
  \label{eq:43}
 t_{c1}=\frac{2\pi}{3}\ \frac{1}{J},\textrm{ and }   t_{c2}=
 \frac{4\pi}{3} \ \frac{1}{J}, 
\end{equation}
the Potts chain behaves as a disordered phase characterized by the
high temperature fp $y^*=1$ with $f(t)$ given by Eq. (\ref{eq:39}).
There is no difference between the open and periodic boundary
conditions because the spins (or, rather, dual spins) behave
independently. For the time interval between A$_1$, A$_2$, given by
Eq. (\ref{eq:43}), the rate function is characteristic of the fp
$y^*=-2$. (Eq. (\ref{eq:81})).  There is a periodicity, and the
transitions take place at $t_{c1}+ 2\pi n/J$, and $t_{c2}+ 2\pi n/J,$
for integer $n$.  These transitions were pointed out in Ref.
\cite{ks17}. There is a phase that has no analog in the thermodynamic
system.

\subsection{The Nature of  transitions}
\label{sec:nature-transitions}
To explore the nature of the phase transitions, we first note that
A$_1$ and A$_2$ form a period-2 cycle so that each one is a repulsive
fixed point of $R^{(2)}$ with multiplier $\lambda=2^2$, (see Table
\ref{tab:1} of Appendix \ref{sec:peri-orbits-potts}). {{The RG
    transformation $R^{(2)}$ requires a decimation of exactly four
    bonds, which results in a length rescaling factor of $s=4$.
    Therefore, in this one-dimensional ($d=1$) case, $\lambda =s^d$.
    This relationship between the fixed point's $\lambda$ and the
    rescaling factor $s$ is a characteristic of a ``discontinuous''
    fixed point for a first-order transition \cite{nienhuis}. }}

{{We now show that $f(y)$ has slope discontinuities at the transition
 points.}} Let us assume
power-law singularities around such a ``discontinuous'' fp, generically denoted by
$y_c$, of the form
\begin{subequations}
\begin{equation}
  \label{eq:55}
  f(y)\sim f_0\, |y-y_c|^{2-\alpha},
\end{equation}
using  the standard notation  for critical exponents \cite{huang,smb-serc}.

The relation for the free energy as given by Eq. (\ref{eq:25}) can be
written in terms of $R^{(2)}$ as
\begin{equation}
  \label{eq:54}
  f_n(y)=\frac{1}{2^2}f_{n-2}(R^{(2)} (y)) + g_2(y),
\end{equation}
where $g_2(y)=\frac{1}{2}\ln c(y) + \frac{1}{2^2} \ln c(R(y))$ is an
analytic function around each of the fps ($y=R^{(2)}(y)$)
$$y_{c1}=e^{i2\pi/3} \textrm{ and } y_{c2}=e^{i4\pi/3}.$$

Expanding around $y_c$,
\begin{equation}
  \label{eq:56}
  R^{(2)}(y)=\lambda\, (y-y_c), \textrm{ with } \lambda=2^2,
\end{equation}
we rewrite Eq. (\ref{eq:54}) for large $n$ (when   $f_n(y)$ is
independent of $n$) as
\begin{equation}
  \label{eq:57}
  f(y)\approx \frac{1}{2^2} f(R^{(2)}(y)),
\end{equation}
which implies, for the singular part,
\begin{equation}
  \label{eq:59}
    |y-y_c|^{2-\alpha}= \frac{\lambda^{2-\alpha}}{2^2} |y-y_c|^{2-\alpha},
  \end{equation}
so that 
\begin{equation}
  \label{eq:58}
  \frac{\lambda^{2-\alpha}}{2^2}=1, {\textrm{ or, } } \alpha=1,
\end{equation}
\end{subequations}
i.e., $f(y)$ behaves linearly near $y_c$.  Therefore, at both the
points $A_{1}, A_{2}$, there are slope discontinuities in $f(t)$ as
seen in Fig. \ref{fig:fvstoned}.  The transitions in time are akin to
first-order transitions.  Surprisingly, the transitions are described
by a cycle of period-2 or by fps of bigger blocks for RG, rather
unusual when compared with thermal phase transitions.

\subsection{Open versus periodic boundary conditions}
\label{sec:open-versus-periodic-1}

The rate function $f(t)$ for the Potts chain depends on the boundary
condition (bc) imposed on the chain, open or periodic {{(see Sec.
    \ref{sec:open-bound-cond}-C)}}.  This difference contradicts the
typical behavior observed in thermodynamic systems' bulk behavior.
The open chain, with free boundary spins, does not exhibit DQPT, while
the periodic chain does, despite having identical RG flow equations
and fixed points.

The distinction arises from the remainder or boundary term, (see Sec. \ref{sec:contr-with-rg})
\begin{equation}
  \label{eq:16}
 b_n= 2^{-n} \ln Z_1(y^{(n)}), 
\end{equation}
in the RG computation of the rate function, Eq. (\ref{eq:26}).
{{The behaviour of  $b_n$ for $n\to\infty$ is analyzed below for the
thermal and the quantum cases. }}

\subsubsection{Thermal case}
\label{sec:paragr-case}

\label{sec:thermal-case}

The boundary term, Eq. (\ref{eq:16}), in the $n\to\infty$ limit
requires the fp value of $y$.  For the thermal problem (real $y$), the
partition function for the finite system is never zero and, therefore,
$2^{-n} \ln Z_1\to 0.$ In this case, the free energy is determined
solely by the sum of the contributions from the RG factors
$g(y^{(j)})$ (Eq. (\ref{eq:26})), and {\it the summation is independent of
the boundary conditions.}

\deltaf

\subsubsection{Quantum case}
\label{sec:quantum-case}
\begin{subequations}

We now consider Eq. (\ref{eq:16}) for the quantum case.  
For large $n$, $y^{(n)}$ approaches one of the two attractive fixed
points, $y^*=1,$ or $-2$ (for the Potts chain), both of which belong
to the Fatou set.

We can categorize the possibilities into two groups. Typically,
the Julia set is comprised of the zeros of the partition function,
while the Fatou set represents the different phases of the system.
Nevertheless, there is the second scenario when the zeros are the
attractive fixed points within the Fatou set. These two cases form the
basis of the boundary-condition dependent dichotomy.
The results (proved below) can be stated as a mathematical mechanism
as follows.
\begin{thm*}[Boundary]
  {\em If an attractive fixed point of the RG transformation, which is
    a member of the Fatou set, coincides with the zero of the
    partition function, the boundary contributions become as
    significant as the bulk rate function. If the singularities cancel
    out exactly, the phase linked to that fixed point would not
    appear.}
\end{thm*}
We establish this theorem under various boundary conditions.

\paragraph{Periodic chain:}
\label{sec:periodic-case}

For the periodic case, the remainder term is determined by
$Z_{\Delta}(y^{(n)})$ (the first generation lattice), which approaches
a finite non-zero number so that $b_n\to 0$ as $y\to y^*.$ This is the
generally expected result, and the RG flow determines the nature of $f(t)$.
{\it{ A DQPT ensues.}}

\paragraph{Open chain}
\label{sec:open-chain}

For the open boundary case, we note that $Z_1$ approaches a constant
when $y^{(n)}\to 1$, and the remainder term vanishes ($b_n\to 0$) in
the limit.  In contrast, over the region characterized by $y^{(n)}\to
y^*= -2,$ extra care is needed because $Z_1=3(y^{(n)}+2)\to 0.$

Defining $z_n=3(y^{(n)}+2),$ the recursion relations for $z_n$ and
$b_n$ can be written with the help of the RG equation $y'=R(y)$ as
\begin{align}
  z_{n+1}&=3(R(y)+2)=\frac{z_n^2}{2z_n-9},  \label{eq:17}\\
  b_{n+1}&=b_n-\frac{1}{2^{n+1}}\ln (2z_n-9).\label{eq:18}
\end{align}
These relations allow us to evaluate 
the boundary-dependent remainder term, $b_{\rm open}(y)$ as 
\begin{equation}
  \label{eq:82}
  b_{\rm open}(y)\equiv\lim_{n\to\infty} b_n =\left\{\begin{array}{ll}
                           0, & {\textrm{ for }} y^*=1,\\
                           {\textrm{nonzero}}, & {\textrm{ for }} y^*=-2.
                           \end{array} \right.
\end{equation}
It follows that  $b_{\rm open}(y)$ is singular at $A_1,A_2,$ the transition
points of Fig. \ref{fig:fvstoned}.

Numerically evaluated $b_{\rm open}(t)$ is shown in Fig.
\ref{fig:deltaf} and is compared with the following ansatz,
\begin{equation}
  \label{eq:83}
 b_{\rm open}(t)= \frac{1}{2} \ln \frac{2-2\cos(Jt)}{5+4\cos(Jt)},\quad \frac{2\pi}{3}<Jt<\frac{4\pi}{3}.  
\end{equation}
\end{subequations}
The exact cancellation of the singularities of $f(t)$ of the periodic
chain (the bulk contribution) by the boundary terms gives us the analytic
behaviour of the open chain of Fig. \ref{fig:b1ft}.  There is no DQPT in this case.

The above arguments predict that there will be no DQPT if one boundary
spin is kept fixed while the other end is free.  Instead, if both the
boundary spins are held fixed, there will be DQPT. A few cases are
listed in Table \ref{tab:bcdqpt}.  These predictions are verified in
the next section by the transfer matrix approach \cite{ks17}.

\tablebc

\subsection{Intermediate phase and Transfer matrix}
\label{sec:interm-phase-transf}

The rate functions are evaluated so far with the help of the series of Eq.
(\ref{eq:26}).  To analyze the intermediate behaviour, we utilize the
transfer matrix approach\cite{huang}.  The partition function for a chain of $B$
bonds can be expressed as a $3\times 3$ matrix which can be
constructed by repeated multiplication of $3\times 3$ symmetric
transfer matrices
\begin{equation}
  \label{eq:47} 
    {\mathbb{M}}=\left(\begin{array}{ccc}
    y&1&1\\
    1&y&1\\
    1&1&y
    \end{array}\right),\,
\end{equation}
so that after $B$ steps, the partition function with the first spin in
state $\alpha_j$ and the last spin in state $\alpha_k$ ($j,k=1,2,3$)
is given by
\begin{equation}
  \label{eq:48}
  Z(y|\alpha_j\alpha_k)=  {\mathbb{M}}^B\big|_{jk}.
\end{equation}
For the periodic boundary conditions, we identify the first and the
last spins, and therefore, the total partition function is
\begin{equation}
  \label{eq:49}
  Z_{\rm pbc}(y)= \textrm{Tr } \mathbb{M}^B.
\end{equation}
For the open boundary conditions, the first and the last spins are
independent, and the partition function involves a sum over all the 9
possible combinations, i.e., the sum of all the terms of the $Z$
matrix so that
\begin{equation}
  \label{eq:50}
  Z_{\rm open}(y)= \textrm{Tr } \mathbb{M}^B + 6  Z(y|\alpha_j\alpha_k), (j\neq k).
\end{equation}
These extra off-diagonal terms do not contribute to the thermodynamic
behaviour but are vital for the quantum problem.

The eigenvalues of $\mathbb{M}$ are \cite{perron}
\begin{subequations}
\begin{equation}
\Lambda_1=y+2,\ \Lambda_2=y-1\label{eq:61}
\end{equation}
where $\Lambda_2$ is doubly degenerate.
Thanks to the symmetric form of
$\mathbb{M}$, the eigenvector of $\Lambda_1$ is $(1,1,1)/\sqrt{3}$,
while the other two can be chosen as $(-1,0,1)/\sqrt{2},
(-1,2,-1)/\sqrt{6}$. The eigenvectors are all real, orthogonal, and
independent of $y$, allowing us to write \cite{craven}
\begin{align}
{\mathbb{M}}^B &= \frac{1}{3^B} \left(\begin{array}{ccc}
                    a&b&b\\
                    b&a&b\\
                    b&b&a
    \end{array}\right),\,  {\textrm{ with }} \left\{\begin{array}{cc}
                                                  a &=\Lambda_1^B+2\Lambda_2^B,\\
                                                  b &=\Lambda_1^B-\Lambda_2^B
                                                   \end{array}\right.  \label{eq:52} \\
\intertext{so that} 
 Z_{pbc}(y)&=\frac{1}{3^B}(\Lambda_1^B+2\Lambda_2^B)\nonumber\\
& \underset{B\to\infty}{=} \left\{\begin{array}{cc}
                              (\Lambda_1/3)^B & {\textrm{if }} |\Lambda_1|>|\Lambda_2|\\
                              (\Lambda_2/3)^B & {\textrm{if }} |\Lambda_1|<|\Lambda_2|
                    \end{array}\right.  \label{eq:53}
\end{align}
For real $y$, $1\leq y<\infty$, we have $\Lambda_1\geq\Lambda_2$,
equality only for $y\to\infty$, and the free energy of the Potts chain
per bond is given by $-k_BT \ln \Lambda_1$ for all temperatures.
However, for the quantum problem with $y=\exp(iJt)$, there is a
crossing of the eigenvalues with degeneracy occurring on the unit
circle $|y|=1$. When $|\Lambda_2| >|\Lambda_1|$, the rate function per
bond is given by
\begin{equation}
  \label{eq:81}
f(t)= -{\textrm{ Re }}\ln (\Lambda_2/3) = -\frac{1}{2} \ln [(2-2 \cos Jt)/9],   
\end{equation}
as shown in Fig. \ref{fig:fvstoned}. This form is observed for
$t_{c1}\leq t\leq t_{c2}$, the range in which $y$ flows to the fp
$y^*=-2$. There is no analog of this phase in the one-dimensional
magnet.

Since the phase transition occurs at the point of degeneracy of the
distinct eigenvalues, a diverging length scale appears close to the
transition point as
\begin{equation}
  \label{eq:60}
  \xi=\left|\ln \left|\frac{\Lambda_1}{\Lambda_2}\right|\, \right |^{-1},
\end{equation}
such that $\xi\to\infty$ as $t\to t_{c1}\pm$ or $t\to t_{c2}\pm$.
This diverging length scale is to be interpreted as a finite-size
length scale such that for a finite system, deviations from the sharp
transition is seen for $B<\xi(t)$.

By using the known eigenvalues Eq.(\ref{eq:61}), the divergence of
$\xi$ is given by
\begin{equation}
  \label{eq:62}
 \xi \sim |t-t_c|^{-1},{\textrm{ both for } } t_c=t_{c1} {\textrm{ and }}t_{c2},
\end{equation}
and to see the sharpness of the transition, we need to respect $B\gg
\xi$.  Or, finite $B$ data can be made to collapse on a master curve
(data collapse) \cite{smbseno} if $B/\xi$ is used as a scaling
behaviour.  Therefore, for a chain of $B$ bonds, the rate function
$f_B(t)$ per bond satisfies a scaling form
\begin{equation}
  \label{eq:63}
  f_B(t)\equiv f(t_c)+ B\; {\cal{F}}\left(\frac{B}{\xi(t)}\right),
\end{equation}
where the scaling function 
\begin{equation}
  \label{eq:64}
{\cal{F}}(x) \approx f_{\pm}\,  \frac{1}{x}, {\textrm{ as } } t\to t_c\pm. 
\end{equation}
with $f_{\pm}$ determining the jump in the slope at $t_c$.  It is
straightforward to verify such a data collapse  from numerics.

The same scale $\xi$ determines the spatial exponential  decay of spin-spin
correlations in both the phases characterized by $y^*=1$ and $y^*=-2$,
indicating that these phases are paramagnetic.

\subsubsection{Importance of freedom}
\label{sec:importance-freedom}
The RG analysis can be compared with the transfer matrix results.  In
the open chain case, Eq. (\ref{eq:50}) shows a perfect cancellation of
the contributions from $\Lambda_2$. However, changes in the free
conditions of the boundary spins would tilt the behaviour towards the
DQPT case. There is no DQPT if we keep one boundary spin fixed at,
say, state-1, so that the partition function involves a summation of
the elements across one row of the matrix in Eq. (\ref{eq:52}). On the
other hand, if the two boundary spins are kept fixed, only one element
of the matrix is needed, and {\it there is DQPT}.  All the entries of
Table \ref{tab:bcdqpt} can be verified with the help of Eq.
(\ref{eq:52}).
\end{subequations}

\section{higher dimensions $(d>1)$}
\label{sec:higher-dimensions-d1}

Dimensionality ($d$) is important in determining the nature of phase
transitions in classical and quantum systems.  We expect $d$ to be
important for DQPT, too, thanks to the intimate connection between
DQPT and classical phase transitions. For example, in classical
systems with finite range interactions and discrete symmetry, a
transition occurs for $d>1$ (Peierls' argument) but not in $d=1$
(Landau's argument) \cite{huang}.  From the perspective of the
renormalization group, these results mean that the zero-temperature
fixed point ($y^*=\infty$), which is unstable for $d=1$ (see Eq.
(\ref{eq:30})), must become stable for $d>1.$ An immediate consequence
is that the Julia set, relevant for DQPT, is compact, i.e., restricted
to the finite complex plane.

The following discussion delves into the RG transformation and its
fps. We also explore the nature of the Julia set, a fractal, in Sec.
\ref{sec:nature-julia-set}. Furthermore, we examine the DQPTs
corresponding to the Julia set's repulsive fps on the unit circle.
These fps are analogous to the discontinuous fp for first-order
transitions. The details of the transition behaviour can be found in
Sec. \ref{sec:transition-points}, where the rate function $f(t)$ is
also obtained.

Let us consider a two-dimensional hierarchical lattice, i.e., with b=2
(Sec.\ref{sec:hier-latt}). The renormalization transformation is now
\begin{equation}
  \label{eq:65}
  R_2(y)= R(y)^2= \left(\frac{y^2+2}{2y+1}\right)^2,
\end{equation}
which is different from $R^{(2)}$ of Eq.(\ref{eq:56}).

The complex dynamics protocol reveals that $R_2(y)$ has a degree of
four, leading to five fixed points, including the point at infinity.
At low temperatures, the ordered state remains stable, making
$y=\infty$ a stable fp with a basin of attraction near the point at
infinity in the extended complex-$y$ plane. This region belongs to the
Fatou set.

The fixed points and their multipliers  are
\begin{subequations}
\label{eq:68}
\begin{align}
  y^*=1,&\quad \lambda=0,\\
 y^*=4, &\quad \lambda=16/9,\\ 
 y^*=e^{\pm i 2\pi/3},& \quad \lambda=4 e^{\mp i 2\pi/3},\quad 
   |\lambda|=4 ,\\
 y^*=\infty,&\quad \lambda=0.
\end{align}
\end{subequations}
Of these $y^*=1,$ and $\infty$ are the two stable (super-attractive,
because the multipliers $=0$) fixed points representing the para or
high-temperature phase and the ordered phase, respectively, with
$y^*=4$ as the thermodynamic critical point (``Curie point'')
separating the two phases.  The remaining two fixed points are on the
unit circle and are repulsive or unstable. These are relevant for
DQPT.

\subsection{Nature of the Julia set}
\label{sec:nature-julia-set}

A notable difference between $b=2$ and the one-dimensional $b=1$ case
is the behaviour of the $y=\infty$ fp, which is now a super-attractive
fp, with its own basin of attraction.  The Julia set, as pointed out
earlier, contains an infinite number of points that do not flow to the
stable fps \cite{saupe}, (see also Sec. \ref{sec:julia-set-fatou}).

\juliabtwo

The RG map has critical points (where the first derivative is either
zero or infinity) at
\begin{itemize}
\item  $y=1,\infty$ (fp) 
\item  $y=-2, \pm i \sqrt{2},$ (not fps)
\item  $y=-1/2$ (a pole).  
\end{itemize}
Of these $y=-2, \pm i \sqrt{2}$ are
in the Julia set, as repeated iterations give \cite{commjulia}
\begin{equation}
-2 \underset{R_2}{\longrightarrow} 4\underset{R_2}{\longrightarrow} 4,
{{\textrm{ and }}}
 \pm i \sqrt{2}\underset{R_2}{\longrightarrow}  0 \underset{R_2}{\longrightarrow} 4.\label{eq:72}
\end{equation}

The Julia set for $R_2$ is shown in Fig. \ref{fig:b2julia}. The colour
code indicates the number of iterations ($<20$) required to reach the
stable fps ($y^*=1$ or $\infty$).  The unstable fp at $y^*=4$ where
the zeros pinch the real axis is the classical critical point of the
$q=3$-Potts model for $b=2$. The two complex fps are $A_1, A_2$ of the
b=1 case, Fig. \ref{fig:onedcircle}. A closer view of the area around
$A_1$ is shown in Figure \ref{fig:b2julia}(b), while Figure
\ref{fig:b2julia}(c) provides a display of the region close to the
imaginary axis on the first quadrant.  (See Appendix
\ref{sec:julia-set-b=2} for the Julia set only.)
\subsection{Transition points}
\label{sec:transition-points}

The intersection points of the Julia set with the unit circle (red
circle in Fig. \ref{fig:b2julia}) are the transition points for DQPT,
while the intersection with the positive real axis gives the thermal
critical point. The absence of isolated points in the Julia set
ensures that the zeros of the partition function for $b=2$ are not
isolated and, therefore, stand for points of phase transitions.
Suppose a continuum limit is taken to describe the zeros by a line or
surface density of ``charges''. In that case, the nature of the
transition is completely determined by the behaviour of the density at
the intersection point. On the other hand, if the intersection point
is an unstable fp, the RG procedure would suffice to determine the
transition behaviour.

The unstable fp at $y^*=4$ describes the thermal criticality of the
Potts model. The temperature derivative of the specific heat is
divergent (see Eq. (\ref{eq:58}) ) with
\begin{equation}
  \label{eq:66}
\alpha=2- \frac{\ln 4}{\ln \lambda}=2- \frac{\ln 4}{\ln (16/9)} =-0.4094... , 
\end{equation}
where $\alpha$ is one of the critical exponents to describe the
symmetry breaking of this particular two-dimensional Potts model.

The two additional  fixed points on the unit circle, 
\begin{equation}
y^*=e^{i2\pi/3},{\textrm{ and }}  e^{i4\pi/3}=e^{-i2\pi/3},   
\end{equation}
correspond to $A_1$ and $A_2$ in Fig. \ref{fig:onedcircle}. 

In the decimation process of the two-dimensional ($d=2$) lattice, the
length is rescaled by a factor of $s=2$, so that the multiplier can be
written as $\lambda=4= s^d$.  This enables us to identify the two
fixed points $A_1$ and $A_2$ as ``discontinuous'' fixed points, which
signify first-order transitions.

The RG 
flow takes $y$ to $y^*=1$ for the arc $A_1$ to $A_2$ on the right
side, similar to the $d=1$ case.  However, on the left part of the arc,
the flow is to $y=\infty$, indicating that the system is in the
ordered state.  There is a {\em symmetry-breaking transition} in time
at $J t_{c1}=2\pi/3$ and then a symmetry-restoring transition at $J
t_{c2}=4\pi/3$ (periodic para to ferro to para transitions). All these
transitions are described by fixed points different from the
thermodynamic one.

We determine the exponent $\alpha$ at $A_1$ (and, by symmetry,
identical behaviour at $A_2$, though time-reversed) from Eq.
(\ref{eq:59}) with a rescaling factor of $2b=4$
\begin{equation}
  \label{eq:67} 
  \alpha=2- \frac{\ln 2b}{\ln |\lambda|}=1,
\end{equation}
by using the multiplier noted in Eq. (\ref{eq:68}). This signals a
first-order transition (slope-discontinuity of $f(t)$).

\btwoftvst

A gross feature of the quantum quench in the Potts model is
the oscillatory behaviour of para and ferro states with the
symmetry-breaking (and restoring) transitions taking place at times
$t_{c1}, t_{c2}$ given by Eq. (\ref{eq:43}).  Fig.  \ref{fig:b2ft}
shows the Loschmidt rate function obtained by using the sum formula of
Eq.  (\ref{eq:26}). The remainder term does not contribute here.  The
major transition is (Point $A_1$) described by a fp (Eq.
(\ref{eq:67})) and is indicated by the arrow to the peak. The
transitions occur periodically at $t_{c1}+ 2\pi n/J$ and $t_{c2}+2\pi
n/J$. 

In addition to the above gross features, the fractal nature of the
Julia set creates finer structures in the time evolution.  These
structures can be determined by monitoring the flow of $y$ along the
unit circle or by studying the unit circle's intersection with the
Julia set.  Fig. \ref{fig:b2julia} shows that the unit circle passes
through regions of the Julia set, fragmenting the arc of the unit
circle into segments flowing to either $y=1$ or $y=\infty$.  Multiple
para-ferro-para transitions are expected in a cycle, especially in
regions close to $Jt=\pi/2$ and around $A_1$. Similar regions also
exist in the lower half of the unit circle due to symmetry.  Fig.
\ref{fig:b2ft} displays the transition points from the flow of $y$
(represented by yellow lines).  Unlike the major transitions, these
minor transitions are not connected to the fps but, as members of the
Julia set, should be related to some higher-order periodic or chaotic
orbit. Further exploration of these minor transitions is imperative.

Successive iterations of the RG transformation are equivalent to
choosing more spins for decimation .  $R_b(y)$ ($R$ or $R_2(y)$ for
$b=1,2$) is obtained by decimating the motifs one at a time, while
$R_b^{(n)}$ would be an RG scheme of a larger step.  Therefore,
periodic orbits are equivalent to fps of $R_b^{(n)}$ of periodicity
$n$, though it indicates a subtle connection between space and time
transformations, respecting time reversibility.  Note that in these
one-parameter systems, the thermal problem cannot display any
periodicity, which is only feasible in the complex plane for the
quantum problem.

Both fps, $A_1$ and $A_2$, can be considered as the fps of $R^{(2)}$
in the one-dimensional scenario. Nevertheless, when seen as an
iteration of $R(y)$, $A_1 \leftrightarrow A_2$. Decimation in space
leads to a transformation of time, causing $A_1$ to become $A_2$. This
transformation can be interpreted as the system being in the future or
the past ($e^{i Jt}=e^{i (Jt\pm 2\pi)}$), a consequence of the time
reversibility of the process.

\section{Conclusion}
\label{sec:conclusion}
In this paper, we investigated dynamical quantum phase transitions
(DQPT) in the quantum 3-state Potts model (Eq. (\ref{eq:7})) using the
iterated map formulation (complex dynamics) for exact renormalization
group (RG) transformation. We focussed on one and two-dimensional
lattices built hierarchically as examples of the exact formulation.
The DQPT occurs during the time evolution of the interacting system,
and here, we started from a state of uniform superposition of all
states that a large transverse field can produce.  The time evolution
here is unitary and the time scale (restoring $\hbar$) is $tJ/\hbar$.
This means that there is no classical limit as $\hbar\to 0$, and the
reported phenomena are inherently quantum in nature.

The RG transformation of the Boltzmann factor $y$ is a rational
function that divides the complex-$y$ plane into (a) the Fatou set of
points that flow to the attractive fixed points (possible phases of
the system), and (b) the Julia set consisting of the repulsive fixed
points (possible transition points) and repulsive periodic orbits.
The Julia set is the boundary of the Fatou set and plays a vital role
in the formulation because of its connection to the zeros of the
partition function. The description of DQPT requires an analysis of
the Julia set, and the RG flows.  The rate function determining the
Loschmidt echo and the thermodynamic free energy are determined
exactly (or numerically exactly) from the RG flows.

From the studies of the open and periodic Potts chains (and other
cases), we conclude that the coincidence of an attractive fixed point
of the renormalization group transformation with the zero of the
partition function makes boundary contributions to the rate function
on par with the bulk, to the extent of even annulling DQPT.  For the
one-dimensional case, the DQPTs are between two paramagnetic phases
described by two different stable (i.e., attractive) fixed points. The
transitions are first-order and are characterized by a periodic cycle
of order 2.

For the two-dimensional models, we find alternating symmetry-breaking
and restoring transitions between the para and ferro phases.  The
transitions are first-order, described by RG fixed points related by
time-reversibility.  {{To be noted here that the fixed point description
for the first-order transitions are consistent with the discontinuous
fixed points known for equilibrium phase transitions.}} 

We may contrast the DQPT results with the thermal problem.
One-dimensional Potts chain does not undergo any thermal phase
transition, while, in two dimensions, there is a symmetry-breaking
critical point (Curie point).

It has not escaped our notice that the first-order dynamic quantum
phase transitions we observed in the Potts system of discrete
symmetry-breaking immediately suggests the possibility of interfaces
separating the two coexisting phases. The nature of the interface in
the quantum system remains to be elucidated.

Even for approximate real space RG for the Boltzmann factor, the
transformation for $y$ can be represented in terms of rational
functions, and then the theory used here can be applicable.  The
complex dynamics formulation can be extended to many other systems
including models in higher ($>2$) dimensions, and also for more than
one variables.

\begin{acknowledgments}
  The author thanks Jaya Maji for discussions on python and
  mathematica programs used in this paper.  The author thanks Oleg
  Evnin and other organizers of the `5th Bangkok Workshop on Discrete
  Geometry, Dynamics and Statistics', January 2023, at the physics
  department of Chulalongkorn University, Bangkok \cite{bangkok},
  where some of the results of this paper were presented.
\end{acknowledgments}

\vfill

\newpage

\appendix

{{
\section{RG equations}
\label{sec:rg-equations}

Some details of the RG procedure of  Sec. \ref{sec:impl-rg-compl}
are given here.  The RG steps
are indicated schematically in Fig. \ref{fig:rgappend}. 

 Note that a bond with two Potts spin in the same
state has a weight of $y=e^{\beta J}$, while a bond with two different
spins has a weight of $1$.    The decimation process
involves removing (by summing over) the spins at the red square
sites within each pink ellipse, while keeping the spins at the blue
disk sites fixed. The new lattice is obtained by joining
the blue sites with bonds.  These are Steps 1 and 2 of Sec. \ref{sec:impl-rg-compl}.
The blue spins are now coupled by a renormalized interaction or a
renormalized weight $y'$. See Fig. \ref{fig:rgappend}. 

The renormalization procedure preserves the physical properties, such as
the partition function.  The renoralized Hamiltonian is similar to the
original one upto an additive constant.  The additive constant in
energy is the multiplicative factor $c$ in the RG steps. 
The two unknows $y'$ and $c$ are then determined by equating the
partition functions.

\begin{figure}[htbp]
  \centering
  \includegraphics[width=\linewidth]{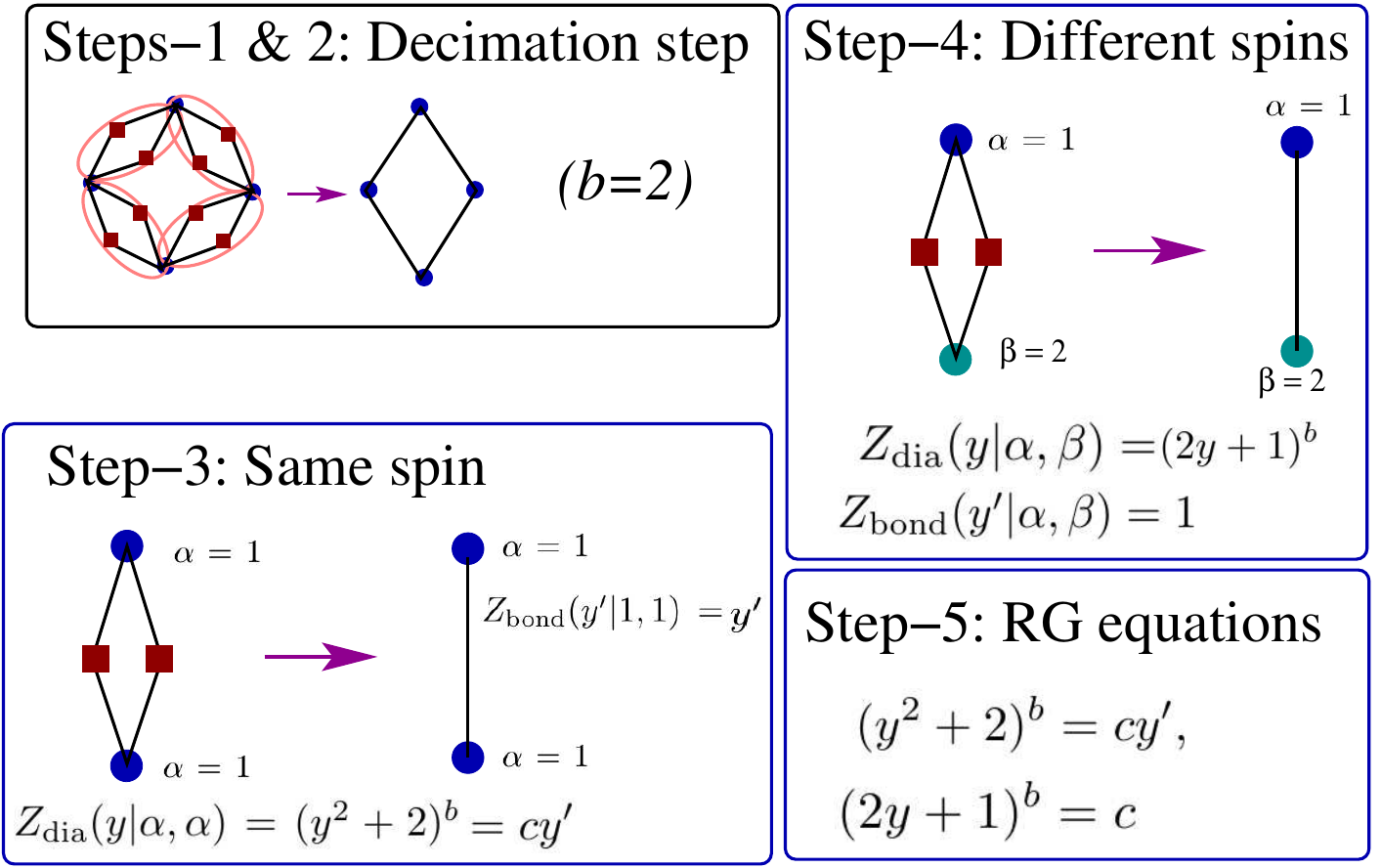}
  \caption{The steps of RG transformation given in
    Sec. \ref{sec:impl-rg-compl} are shown schematically.  The
    decimation is done by summing over the spins at the re square
    sites.
        }%
  \label{fig:rgappend}
\end{figure}
}}

\onecolumngrid

\section{Recursion relations for the partition function}
\label{sec:recurs-relat-part}

For small lattices, the Loschmidt amplitudes can be computed by
constructing the recursion relations as the lattice is built up. By
using the permutation symmetry of the spin orientations, the relations
are written as
\begin{subequations}
\begin{align}
Z_n(y|\alpha,\alpha) &= \left[Z_{n-1}(y|\alpha,\alpha)^2
\right.
       +\left. (q-1) Z_{n-1}(y|\alpha,\beta)^2\right]^b, 
                       (\beta\neq\alpha),\label{eq:27}\\
Z_n(y|\alpha,\beta) &=\left[2 Z_{n-1}(y|\alpha,\alpha)  Z_{n-1}(y|\alpha,\beta)\right.
+
\left.Z_{n-1}(y|\alpha,\gamma) Z_{n-1}(y|\gamma,\beta)\right]^b,
(\beta\neq\alpha, \gamma\neq\alpha,\gamma\neq\beta), \label{eq:28}\\
Z_{n}(y)&= q Z_n(y|\alpha,\alpha)+ q(q-1) Z_n(y|\alpha,\beta),
{\textrm{ any }} \alpha\neq\beta.
\end{align}
\end{subequations}

\section{Periodic orbits for the Potts chain}
\label{sec:peri-orbits-potts}
For the one dimensional chain, there are no unstable fixed points, except the 
point at infinity.  However,  the Julia set contains many periodic orbits. 
The orbits of periodicity $2$ to $5$ are listed in Table II.

\begin{table*}[htbp]
\begin{ruledtabular}
\begin{tabular}{lll}
period& points $(\eta): z=-0.5+i \eta$ & $\lambda$\\
\hline
2 &$\pm \sin 2\pi/3$ & $2^2$\\
\hline
3 & ${\rm P}_3=\{-3.11478, -1.19621,0.342365\}, $ & $2^3$\\
  & $-{\rm P}_3$& \\
\hline
4 & ${\rm P}_{4,1}=\{ -7.05695, -3.36906, -1.35061, 0.157656\}, $& $2^4$\\
  & $-{\rm P}_{4,1},$ & \\
  & ${\rm P}_{4,2}=\{-2.06457, -0.48738, 2.06457, 0.48738\}, $ & \\
\hline
5  & ${\rm P}_{5,1}=\{-14.7507, -7.29908,-3.49541, -1.42586, +0.0760714\},$ & $2^5$\\
  & $-{\rm P}_{5,1},$ & \\
  & $-{\rm P}_{5,2}=\{-1.16109, 0.388377, -2.70248, -0.934957, 0.735785\},$ & \\
  & ${\rm P}_{5,3}=\{-0.555539, 1.74729, 0.229792, -4.78085, -2.15511\},$ & \\
  & $-{\rm P}_{5,3}$ & \\
\end{tabular}
\end{ruledtabular}
\caption{For the one-dimensional Potts model, the periodic orbits and
  their multipliers. The points are on the line $z=-0.5+i \eta,$ and
  only the $\eta$ values are tabulated. } \label{tab:1}
\end{table*}

\vfill

\pagebreak

\section{Julia set for the $b=2$ Potts model}
\label{sec:julia-set-b=2}
\begin{figure*}[htbp]
  \centering
  \includegraphics[width=0.6\linewidth,trim=10pt 10pt 10pt 10pt]{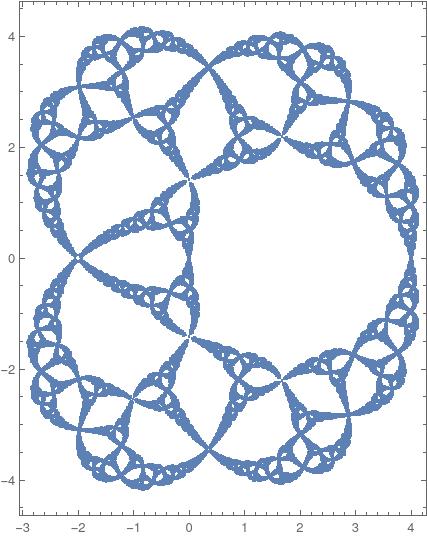}
  \caption{The Julia set for the 3-state Potts model on the $b=2$
    diamond hierarchical lattice, obtained   from {\small
      MATHEMATICA } \cite{mathematica}.
        }%
  \label{fig:b2math}
\end{figure*}

The Julia set for the $b=2$ case is shown in Fig. \ref{fig:b2math}.
The unstable fixed point $y^*=4$ of $R_2(y)$ from Eq. (\ref{eq:65})
belongs to the set, as do the preimages $y=0,-2$ such that $R_2(y)=4$.
The color code in Fig.\ref{fig:b2julia} is based on the number of
iterations needed for convergence.

\vfill

\pagebreak

\twocolumngrid


\begin{thebibliography}{99}
\bibitem{commdiv} More properly, the transition time is the time
  around which there is no  Taylor series
  expansionin $t$  of any relevant quantity like $\ln P(t)$.
\bibitem{diam} The diamond motif is not the only possibility.  One may consider many
other possible motifs \cite{hier}.
\bibitem{commfo}  Iterated maps are also denoted by $R^{on}$ for the
  $n$th iterate.
\bibitem{commdist} If required, the distance between two points $z_1$
  and $z_2$ in the extended plane $\widehat{\cal{C}}$ can be taken as
  the arc-distance on the Riemann sphere or the Euclidean distance of
  the chord in 3-d.  A metric is useful for trajectories going to
  $\infty$.
\bibitem{equicont} More
  formally, if, for a point $z_0$, for every $\epsilon>0$ we can find
  a $\delta>0$ such that $|R^{(n)}(z_0)-R^{(n)}(z)|<\delta,$ for all
  $n$, implies $|z-z_0|<\epsilon$, then the trajectories are
  equicontinuous.
\bibitem{qm2} The $q$-state Potts model has a fixed point at $-(q-1)$.
  In Eq. (\ref{eq:14}), given a value of $\alpha_j$, only one state of
  the neighboring spin $\alpha_j$ contributes to the energy, while the
  remaining $q-1$ states do not contribute.  See Eq. (\ref{eq:24}).
\bibitem{perron} Note that as $y$ may be complex, the Perron-Forbenius
  theorem is not applicable, unlike the real $y$ case. The largest
  eigenvalue can be degenerate.
\bibitem{commjulia} The preimages of $y=-2$, viz.,
  $-0.3483106997490063... \pm 0.615889967883341... i,{\textrm{ and }}
  0.348310699749006... \pm 3.444317092629529... i,$ have similar
  geometrical or topological neighbourhoods.
\end{thebibliography}
\end{document}